\documentclass[aps,prd,twocolumn,showpacs,floatfix,superscriptaddress]{revtex4}

\usepackage{amsmath}
\usepackage{amssymb}
\usepackage{color}
\usepackage{dcolumn}
\usepackage{graphicx}
\usepackage{multirow}
\usepackage{rotating}
\usepackage{afterpage}
\usepackage{enumerate}

\newcommand{\pt}  {\mbox{$p_T$}}
\newcommand{\dzero}     {D0}
\newcommand{\ttbar}     {\ensuremath{t\bar{t}}}

\newcommand{\ppbar}     {\ensuremath{p\bar{p}}}
\newcommand{\qqbar}     {\ensuremath{q\bar{q}}}
\newcommand{\herwig}    {\textsc{herwig}}
\newcommand{\pythia}    {\textsc{pythia}}
\newcommand{\alpgen}    {\textsc{alpgen}}

\newcommand{\geant}     {\textsc{geant}}
\newcommand{\mcatnlo}   {\textsc{mc@nlo}}

\newcommand{\met}       {\mbox{$\not\!\!E_T$}}

\newcommand{\metsig}    {\mbox{${\cal S}$$(\met)$}}
\newcommand{\Z}         {\mbox{$Z/\gamma^\star$}}

\newcommand{\etadet} 		{\ensuremath{\eta_{\text{det}}}}

\newcommand{\All}       {\ensuremath{A^{\ell\ell}}}
\newcommand{\defAll}    {\ensuremath{\All = \frac{N(\Delta\eta > 0) - N(\Delta\eta < 0)}{N(\Delta\eta > 0) + N(\Delta\eta < 0)}}}
\newcommand{\Alfb}      {\ensuremath{A^{\ell}_{\rm FB}}}
\newcommand{\defAlfb}   {\ensuremath{\Alfb = \frac{N(q \times \eta>0) - N(q \times \eta<0)}{N(q\times \eta>0) + N(q \times \eta<0)}}}

\newcommand{\lumi}      {9.7~fb$^{-1}$}

\begin{document}

\hspace{5.2in} \mbox{FERMILAB-PUB-13/347-E}

\title{Measurement of the asymmetry in angular distributions of leptons produced in dilepton 
{\boldmath $t\bar{t}$} final states in {\boldmath $p\bar{p}$} collisions at $\sqrt{s}=1.96$ TeV}
\vspace*{0.1cm}

%
\affiliation{LAFEX, Centro Brasileiro de Pesquisas F\'{i}sicas, Rio de Janeiro, Brazil}
\affiliation{Universidade do Estado do Rio de Janeiro, Rio de Janeiro, Brazil}
\affiliation{Universidade Federal do ABC, Santo Andr\'e, Brazil}
\affiliation{University of Science and Technology of China, Hefei, People's Republic of China}
\affiliation{Universidad de los Andes, Bogot\'a, Colombia}
\affiliation{Charles University, Faculty of Mathematics and Physics, Center for Particle Physics, Prague, Czech Republic}
\affiliation{Czech Technical University in Prague, Prague, Czech Republic}
\affiliation{Institute of Physics, Academy of Sciences of the Czech Republic, Prague, Czech Republic}
\affiliation{Universidad San Francisco de Quito, Quito, Ecuador}
\affiliation{LPC, Universit\'e Blaise Pascal, CNRS/IN2P3, Clermont, France}
\affiliation{LPSC, Universit\'e Joseph Fourier Grenoble 1, CNRS/IN2P3, Institut National Polytechnique de Grenoble, Grenoble, France}
\affiliation{CPPM, Aix-Marseille Universit\'e, CNRS/IN2P3, Marseille, France}
\affiliation{LAL, Universit\'e Paris-Sud, CNRS/IN2P3, Orsay, France}
\affiliation{LPNHE, Universit\'es Paris VI and VII, CNRS/IN2P3, Paris, France}
\affiliation{CEA, Irfu, SPP, Saclay, France}
\affiliation{IPHC, Universit\'e de Strasbourg, CNRS/IN2P3, Strasbourg, France}
\affiliation{IPNL, Universit\'e Lyon 1, CNRS/IN2P3, Villeurbanne, France and Universit\'e de Lyon, Lyon, France}
\affiliation{III. Physikalisches Institut A, RWTH Aachen University, Aachen, Germany}
\affiliation{Physikalisches Institut, Universit\"at Freiburg, Freiburg, Germany}
\affiliation{II. Physikalisches Institut, Georg-August-Universit\"at G\"ottingen, G\"ottingen, Germany}
\affiliation{Institut f\"ur Physik, Universit\"at Mainz, Mainz, Germany}
\affiliation{Ludwig-Maximilians-Universit\"at M\"unchen, M\"unchen, Germany}
\affiliation{Panjab University, Chandigarh, India}
\affiliation{Delhi University, Delhi, India}
\affiliation{Tata Institute of Fundamental Research, Mumbai, India}
\affiliation{University College Dublin, Dublin, Ireland}
\affiliation{Korea Detector Laboratory, Korea University, Seoul, Korea}
\affiliation{CINVESTAV, Mexico City, Mexico}
\affiliation{Nikhef, Science Park, Amsterdam, the Netherlands}
\affiliation{Radboud University Nijmegen, Nijmegen, the Netherlands}
\affiliation{Joint Institute for Nuclear Research, Dubna, Russia}
\affiliation{Institute for Theoretical and Experimental Physics, Moscow, Russia}
\affiliation{Moscow State University, Moscow, Russia}
\affiliation{Institute for High Energy Physics, Protvino, Russia}
\affiliation{Petersburg Nuclear Physics Institute, St. Petersburg, Russia}
\affiliation{Instituci\'{o} Catalana de Recerca i Estudis Avan\c{c}ats (ICREA) and Institut de F\'{i}sica d'Altes Energies (IFAE), Barcelona, Spain}
\affiliation{Uppsala University, Uppsala, Sweden}
\affiliation{Lancaster University, Lancaster LA1 4YB, United Kingdom}
\affiliation{Imperial College London, London SW7 2AZ, United Kingdom}
\affiliation{The University of Manchester, Manchester M13 9PL, United Kingdom}
\affiliation{University of Arizona, Tucson, Arizona 85721, USA}
\affiliation{University of California Riverside, Riverside, California 92521, USA}
\affiliation{Florida State University, Tallahassee, Florida 32306, USA}
\affiliation{Fermi National Accelerator Laboratory, Batavia, Illinois 60510, USA}
\affiliation{University of Illinois at Chicago, Chicago, Illinois 60607, USA}
\affiliation{Northern Illinois University, DeKalb, Illinois 60115, USA}
\affiliation{Northwestern University, Evanston, Illinois 60208, USA}
\affiliation{Indiana University, Bloomington, Indiana 47405, USA}
\affiliation{Purdue University Calumet, Hammond, Indiana 46323, USA}
\affiliation{University of Notre Dame, Notre Dame, Indiana 46556, USA}
\affiliation{Iowa State University, Ames, Iowa 50011, USA}
\affiliation{University of Kansas, Lawrence, Kansas 66045, USA}
\affiliation{Louisiana Tech University, Ruston, Louisiana 71272, USA}
\affiliation{Northeastern University, Boston, Massachusetts 02115, USA}
\affiliation{University of Michigan, Ann Arbor, Michigan 48109, USA}
\affiliation{Michigan State University, East Lansing, Michigan 48824, USA}
\affiliation{University of Mississippi, University, Mississippi 38677, USA}
\affiliation{University of Nebraska, Lincoln, Nebraska 68588, USA}
\affiliation{Rutgers University, Piscataway, New Jersey 08855, USA}
\affiliation{Princeton University, Princeton, New Jersey 08544, USA}
\affiliation{State University of New York, Buffalo, New York 14260, USA}
\affiliation{University of Rochester, Rochester, New York 14627, USA}
\affiliation{State University of New York, Stony Brook, New York 11794, USA}
\affiliation{Brookhaven National Laboratory, Upton, New York 11973, USA}
\affiliation{Langston University, Langston, Oklahoma 73050, USA}
\affiliation{University of Oklahoma, Norman, Oklahoma 73019, USA}
\affiliation{Oklahoma State University, Stillwater, Oklahoma 74078, USA}
\affiliation{Brown University, Providence, Rhode Island 02912, USA}
\affiliation{University of Texas, Arlington, Texas 76019, USA}
\affiliation{Southern Methodist University, Dallas, Texas 75275, USA}
\affiliation{Rice University, Houston, Texas 77005, USA}
\affiliation{University of Virginia, Charlottesville, Virginia 22904, USA}
\affiliation{University of Washington, Seattle, Washington 98195, USA}
\author{V.M.~Abazov} \affiliation{Joint Institute for Nuclear Research, Dubna, Russia}
\author{B.~Abbott} \affiliation{University of Oklahoma, Norman, Oklahoma 73019, USA}
\author{B.S.~Acharya} \affiliation{Tata Institute of Fundamental Research, Mumbai, India}
\author{M.~Adams} \affiliation{University of Illinois at Chicago, Chicago, Illinois 60607, USA}
\author{T.~Adams} \affiliation{Florida State University, Tallahassee, Florida 32306, USA}
\author{J.P.~Agnew} \affiliation{The University of Manchester, Manchester M13 9PL, United Kingdom}
\author{G.D.~Alexeev} \affiliation{Joint Institute for Nuclear Research, Dubna, Russia}
\author{G.~Alkhazov} \affiliation{Petersburg Nuclear Physics Institute, St. Petersburg, Russia}
\author{A.~Alton$^{a}$} \affiliation{University of Michigan, Ann Arbor, Michigan 48109, USA}
\author{A.~Askew} \affiliation{Florida State University, Tallahassee, Florida 32306, USA}
% ...
%
%
\collaboration{The D0 Collaboration\footnote{with visitors from
$^{a}$Augustana College, Sioux Falls, SD, USA,
$^{b}$The University of Liverpool, Liverpool, UK,
$^{c}$DESY, Hamburg, Germany,
$^{d}$Universidad Michoacana de San Nicolas de Hidalgo, Morelia, Mexico
$^{e}$SLAC, Menlo Park, CA, USA,
$^{f}$University College London, London, UK,
$^{g}$Centro de Investigacion en Computacion - IPN, Mexico City, Mexico,
$^{h}$Universidade Estadual Paulista, S\~ao Paulo, Brazil,
$^{i}$Karlsruher Institut f\"ur Technologie (KIT) - Steinbuch Centre for Computing (SCC),
$^{j}$Office of Science, U.S. Department of Energy, Washington, D.C. 20585, USA and
$^{k}$Laboratoire de Physique Theorique, Orsay, FR.
}} \noaffiliation
\vskip 0.25cm

\author{}
\date{August 30, 2013}


\begin{abstract}
We present measurements of  asymmetries in angular distributions of leptons produced in 
\ttbar\ events in proton-antiproton collisions at the Fermilab Tevatron Collider.
We consider final states where the $W^{\pm}$ bosons from top quark and antiquark decays both decay into $\ell\nu$ ($\ell = e, \mu$) resulting in oppositely charged dilepton final states with accompanying jets. 
Using \lumi\ of integrated luminosity collected with the \dzero\ detector,
we find the asymmetries in lepton pseudorapidity
compatible with predictions based on the standard model. 

\end{abstract}

\maketitle 


The top quark, first observed by the CDF and \dzero\ Collaborations in 1995~\cite{Abe:1995hr,Abachi:1995iq}, is 
the heaviest of all elementary particles.
Because of the large top-quark mass, the measurement of the production and 
decay properties of top quark pairs in proton-antiproton ({\ppbar}) collisions provides
an important test of the standard model of particle physics (SM) that may unveil 
the presence of new phenomena beyond the SM (BSM).

Perturbative quantum chromodynamics (pQCD) at leading order (LO) predicts that top quark-antiquark
(\ttbar) production in quark-antiquark (\qqbar) annihilation in the center of mass frame is 
forward-backward (FB) symmetric in the angular distributions of the $t$ and $\bar{t}$ quarks. 
However, a positive FB asymmetry appears from next-to-leading order (NLO) 
contributions~\cite{Bernreuther:2010ny,Hollik:2011ps,Ahrens:2011uf,Kuhn:2011ri}, such that the top (antitop) quark is 
preferentially emitted in the direction of the incoming quark (antiquark).
Processes beyond the SM can modify the \ttbar\ production asymmetry, 
for example through contributions from axigluons or diquarks~\cite{Ferrario:2008wm,Ferrario:2009bz,Antunano:2007da,Frampton:2009rk,Jung:2009pi,Arhrib:2009hu,Djouadi:2009nb,Alvarez:2011hi,Chen:2010hm,Jung:2010yn,AguilarSaavedra:2011ci,Tavares:2011zg,Barcelo:2011vk}, 
$Z'/W'$ bosons~\cite{Cheung:2009ch,Xiao:2010hm,Cao:2010zb,Jung:2009jz,Cao:2009uz,Barger:2010mw}, supersymmetry~\cite{Bauer:2010iq,Chivukula:2010fk,Dorsner:2009mq}, 
or new scalar particles~\cite{Shu:2009xf,AguilarSaavedra:2011ug}.
The CDF and \dzero\ Collaborations have performed measurements of the \ttbar\ FB asymmetry
in \ttbar\ decaying to $\ell$+jets final states containing jets, and an imbalance in transverse energy (\met), and just one lepton
($\ell=$$e$ or $\mu$) from $W$ decay where the $W$ is coming from $t$ or $\bar{t}$,
based on data corresponding to integrated luminosities of 9.4~fb$^{-1}$~\cite{Aaltonen:2012it} and 5.4~fb$^{-1}$~\cite{Abazov:2011rq}, respectively.
The FB asymmetry reported by the CDF and \dzero\ Collaborations both
differ by more than two standard deviations (SD) from the NLO pQCD 
predictions~\cite{Aaltonen:2012it,Abazov:2011rq}. 

Rather than measuring the FB asymmetry of the top quarks themselves, 
an asymmetry in \ttbar\ events can also be measured from the pseudorapidity~\cite{eta} of the 
single charged lepton in the $\ell$+jets final state. 
In such a measurement, based on
an integrated luminosity of 9.4~fb$^{-1}$ and 5.4~fb$^{-1}$, CDF and \dzero\ found
deviations from NLO pQCD predictions of about three~SD~\cite{Abazov:2011rq} and of 1.7~SD~\cite{2013arXiv1308.1120C}, respectively. 
The \dzero\ Collaboration also reported a similar measurement in dilepton final 
states~\cite{Abazov:2012bfa}, where the $W$ bosons from $t$ and $\bar t$ decays both 
decay into $\ell\nu$ (\mbox{$\ell=$$e$ or $\mu$}), in data corresponding to an integrated 
luminosity of 5.4~fb$^{-1}$. The asymmetry results
reported in Ref.~\cite{Abazov:2012bfa}  combined with the measurement in the $\ell$+jets
final state, reduce the disagreement with the NLO pQCD predictions to 2.2~SD~\cite{Abazov:2012bfa}. 

The results of the ATLAS and CMS Collaborations
based on the difference of top and antitop quark production angles in the $\ell$+jets final states
show good agreement with NLO pQCD expectations in proton-proton collisions at
${\sqrt s} = 7$~TeV~\cite{atlas,cms}.
However, at the LHC, measured asymmetries in top quark angular distributions
are not directly comparable with the values extracted at the Tevatron,
because of the symmetry of the initial proton-proton state at the LHC.
This symmetry at the LHC leads to a weaker sensitivity to the physics process responsible
for the production asymmetry compared to the Tevatron.

In this article, we report a new measurement of the asymmetry in the pseudorapidity distributions of leptons 
produced in \ttbar\ events in the dilepton channel, based on all the
data collected by the \dzero\ Collaboration in Run~II of the Tevatron, 
and we compare our results with the most recent predictions based on the standard model~\cite{Bernreuther:2012sx}.
corresponding to an integrated luminosity of \lumi\ following relevant data quality selection.

We use the two observables $q\times\eta$ and $\Delta\eta$, where 
$q$ and $\eta$ are the charge and pseudorapidity of the lepton, and 
$\Delta\eta = \eta_{\ell^+} -\eta_{\ell^-}$ is the difference in lepton pseudorapidities.
The single-lepton asymmetry \Alfb\ is defined as
\begin{equation}
\defAlfb, 
\label{eq:al}
\end{equation}
where $N$ corresponds to the number of leptons satisfying a given set of selection criteria. 
In this asymmetry, each event contributes twice, once with positive and once with negative
lepton charge. The dilepton asymmetry \All\ is defined as
\begin{equation}
\defAll
\label{eq:all}\ . 
\end{equation}
The \Alfb\ and \All\ asymmetries are highly correlated as we discuss in Sec.~\ref{sec:results}.

\section{The \dzero\ Detector and Object Identification}
\label{sec:det}

The \dzero\ detector~\cite{run2det,Angstadt2010,Abolins2008} has a central tracking system consisting of a
silicon microstrip tracker and a central fiber tracker,
both located within a 2~T superconducting solenoidal
magnet, with designs optimized for tracking and
vertexing at detector pseudorapidities (relative to the center of the \dzero\ detector) of 
$|\etadet|<3$ and $|\etadet|<2.5$, respectively.
A liquid-argon sampling calorimeter has a
central section (CC) covering pseudorapidities $|\etadet|$ up to
$\approx 1.1$, and two end calorimeters (EC) that extend coverage
to $|\etadet|\approx 4.2$, with all three housed in separate
cryostats~\cite{run1det}. An outer muon system, at $|\etadet|<2$,
consists of a layer of tracking detectors and scintillation trigger
counters in front of 1.8~T toroids, followed by two similar layers
after the toroids~\cite{run2muon}.

In the current analysis, we focus on \ttbar\ dilepton final states that contain two isolated charged leptons ($ee$, $e\mu$, or $\mu\mu$),
at least two candidate $b$-quark jets, and significant \met\ attributed to escaping neutrinos. 
Electrons are identified as energy clusters in the calorimeter within a cone of radius ${\cal R}=\sqrt{(\Delta\eta)^2+(\Delta\phi)^2}=0.2$ 
(with $\phi$ the azimuthal angle), that are
consistent in their longitudinal and transverse profiles
with those expected of an electromagnetic shower.
More than 90\% of the energy of an electron candidate must be deposited  in the 
electromagnetic part of the calorimeter.
Electrons are required to be isolated by demanding that les than $20$\% of its energy deposited in an
annulus of $0.2 < {\cal R} < 0.4$ around its direction.
This cluster has to be matched to a track reconstructed in the central tracking system.
We consider electrons in the CC with  $|\etadet| <1.1$ and in the EC with $1.5 < |\etadet| < 2.5$.
Transverse momentum \pt\ of electrons must be greater than $15$ GeV.
In addition, we use an 
electron multivariate discriminant based on tracking 
and calorimeter information, to reject jets misidentified as electrons. 
It has an 75\%--80\% efficiency to select real electrons, 
and a rejection $\approx 96$\% for misidentified jets.

A muon is identified~\cite{muonID} as a segment in at least one layer 
of the  muon system that is
matched to a track reconstructed in the central tracking system. 
Reconstructed muons must have $p_T>15$ GeV and satisfy two isolation criteria.
First, the transverse energy deposited in the calorimeter annulus
around the muon $0.1 < {\cal R} < 0.4$ ($E_T^{\mu,\text{iso}}$)
has to be less than 15\% of the transverse momentum of the muon ($p^{\mu}_T$).
Second, the sum of the transverse momenta of the tracks in a cone of radius ${\cal R}=0.5$ around the muon track 
in the central tracking system ($p_T^{\mu,\text{iso}}$) has to be less than 15\% of $p^{\mu}_T$.

Jets are identified as energy clusters in the electromagnetic and hadronic parts of the calorimeter
reconstructed using an iterative mid-point cone algorithm with radius ${\cal R}=0.5$~\cite{Blazey:2000qt}
and $|\etadet|< 2.5$.
A jet energy scale  correction is determined by calibrating
the energy deposited in the jet cone  
using transverse momentum balance in photon+jet and dijet events.
When a muon track overlaps the jet cone, 
the momentum of that muon is added to the jet \pt, assuming
that the muon originates from the semileptonic decay of a hadron belonging to the jet.
Jets in simulated events are corrected 
for residual differences in energy resolution and energy scale between data and simulation.
These correction factors are measured by comparing data and simulation 
in Drell-Yan (\Z$\to ee$) plus jets events.

We use a multivariate analysis (MVA) to identify jets originating from $b$ quarks~\cite{Abazov:2010ab}. 
The algorithm combines into a single discriminant variable the information from the  impact parameters of tracks and from variables 
that characterize the properties of
secondary vertices within jets using a single discriminant.
Jet candidates for $b$~tagging
are required to have at least two tracks with $p_T>0.5$ GeV originating from the vertex of the $p\bar{p}$ interaction
and to be matched to a jet reconstructed from the tracks. 

The \met\ is reconstructed from the energy deposited in the calorimeter 
cells, and corrections to $p_T$ for leptons and jets are propagated into the \met.
A significance in \met\  [\metsig] 
is defined for each event through a likelihood discriminant constructed from the ratio
of the \met\ to its uncertainty.

\section{Simulated events}
\label{sec:samples}

Monte Carlo (MC) events are processed through a \geant-based~\cite{geant}
simulation of the D0 detector.
To simulate effects from additional overlapping \ppbar\ interactions, ``zero bias'' events 
are selected randomly in collider data and overlaid on the fully simulated MC events.
Residual differences between data and simulation of electron and muon
\pt\ resolution and identification are corrected
by comparing \Z$\to\ell\ell$ events in data and MC, applying tight requirements
on one of the two leptons and using
the other one to measure efficiencies and resolutions.

We use the NLO generator \mcatnlo~3.4~\cite{Frixione:2002ik,Frixione:2008ym},
interfaced with \herwig~6.510~\cite{Corcella:2000bw} for parton
showering and hadronization, to simulate \ttbar\ events.
The main sources of background in the dilepton channel correspond to
$q\bar{q}\to\Z\to\ell\ell$, diboson production ({\sl WW, WZ, ZZ}), and instrumental background.
The instrumental background arises mainly from multijet and $(W \to \ell \nu)$+jets events in which one or two 
jets are misidentified as electrons or where 
muons or electrons originating from the semileptonic decay of a heavy-flavor hadron appear isolated.
This background is evaluated using data, as described in Sec.~\ref{sec:selection}.
\Z\ events are generated with the tree-level
LO matrix element generator \alpgen~v2.11~\cite{Mangano2003}
interfaced with \pythia~6.409~\cite{Sjostrand2006} (D0 modified tune~A~\cite{Affolder2002})
for parton showering and hadronization. 
Diboson events are generated with \pythia.
The \mcatnlo\ generator uses the CTEQ6M1 set of parton distribution functions~(PDFs),
and all other simulated samples are generated using the CTEQ6L1
PDFs~\cite{Nadolsky:2008zw}.
The \Z\ samples are normalized to the next-to-next-to-leading-order cross section 
computed with the {\sc fewz} program~\cite{Gavin:2010az}.
We separately simulate \Z\ accompanied by heavy-flavor quarks ($b\bar{b}$ or $c\bar{c}$) using \alpgen,
and enhance the corresponding LO cross sections by a factor 
estimated from the NLO values computed with the {\sc mcfm} program~\cite{Ellis:2006ar}.
The diboson samples are normalized to the NLO cross section calculated with  {\sc mcfm}.

In addition, we apply a correction to the $\Z$+jets simulation, based on data~\cite{Abazov:2010kn}, to address
small discrepancy in the modeling of $Z$ boson transverse momentum$p_T^Z$ in the simulation.

In $Z$ boson events the asymmetries defined in Eqs.~(\ref{eq:al}) and  (\ref{eq:all}) 
are not well-modeled in the simulation, especially in the $e\mu$ channel for
$Z/\gamma^\star\to \tau\tau\to e\nu\mu\nu$ events.
We therefore apply an additional correction using \pythia~8~\cite{Sjostrand2008}, which correctly takes into account the
tau lepton polarization and spin correlations for the tau decays.
This reweighting is explained in detail in Sec.~\ref{sec:systematics}.

An interesting class of BSM models that can generate a large \ttbar\ forward-backward asymmetry at tree level
arises from the presence of a color-octet vector particle $G_\mu^a$  (the so-called axigluon) with large mass $m_G$ and chiral couplings.
To check the sensitivity of our measurements to such new phenomena, we generate two  
axigluon samples~\cite{Falkowski:2012cu} and pass these events through the full \dzero\ simulation and reconstruction programs.
Model 1 has a right-handed coupling to the SM quarks 
of $0.8 g_s$ (where $g_s= \sqrt{\alpha_s/4\pi}$ is the QCD coupling) and no left-handed coupling. The axigluon mass is set to
0.2~TeV and the width to 50~GeV. 
Model 2 has a right-handed coupling to light SM quarks 
of $-1.5 g_s$, a coupling of $6 g_s$  to the top quark, and no left-handed coupling, with
the axigluon mass and width set to 2~TeV and 670~GeV, respectively.
Table~\ref{tab:axigluons} summarizes
the values of the asymmetry predicted by these two models.
These models are in agreement with experimental constraints (\ttbar\ resonance
searches and dijet production) from the Tevatron and the LHC, 
but in slight tension with the $\ttbar$ production cross section measurements.
\begin{table}[!hbt]
\caption{Asymmetries predicted by \mcatnlo\ and by the two models of axigluons described in the text.
Uncertainties reflect only the statistical MC contributions. All values are given in \%.
\label{tab:axigluons}}
\begin{tabular}{cccc}
\hline\hline
& Model 1 & Model 2 & \mcatnlo \\
\hline
\All\ & 21.3 $\pm$ 0.6 & 11.3 $\pm$ 0.5 & 3.3 $\pm$ 0.1 \\
\Alfb\ & 14.9 $\pm$ 1.0 & \phantom{0}8.9 $\pm$ 0.8 & 2.4 $\pm$ 0.1 \\
\hline\hline
\end{tabular}
\end{table}

\section {Event Selection and Estimation of Instrumental Background}
\label{sec:selection}

We follow the approach developed in Ref.~\cite{Abazov:2011cq} for the event selection, i.e.
using the criteria listed below:
\begin{enumerate}[(i)]
\item
\label{sel:first}
For the $ee$ and $\mu\mu$ channels,
we select events that pass at least one single-lepton trigger, while for the $e\mu$ channel, we 
consider events selected through 
a mixture of single and multilepton triggers and lepton+jet
triggers.
Efficiencies for single electron and muon  triggers are measured using
\Z$\to ee$ or $\Z\to\mu\mu$ data, and found to be $\approx 99$\%  and $\approx 80$\%, respectively, for dilepton signal events. 
For the $e\mu$ channel, the trigger efficiency is $\approx$ 100\%.

\item We require at least one $p\bar{p}$ interaction vertex in the interaction region with $ |z| < 60$ cm, where
$z$ is the coordinate along the beam axis,  and $z=0$ is the center of the detector.
At least three tracks with $p_T>0.5$ GeV must be associated with this vertex.

\item We require at least two isolated leptons with \mbox{$\pt>15$ GeV},
both originating from the same interaction  vertex.
We consider only muons within $|\etadet|<2.0$ and electrons
within $|\etadet|<1.1$ or $1.5<|\etadet|<2.5$.
The two highest-\pt\ leptons in an event must have opposite electric charges.

\item  
\label{sel:incl}
To reduce the background from bremsstrahlung in the $e\mu$ final state, we require the distance 
in ($\eta,\phi$) space between the electron and the muon trajectories 
to be ${\cal R}(e,\mu)>0.3$.

\item 
\label{sel:jets}
In the $ee$ and $\mu\mu$ channels, we require at least two jets with $\pt>20$~GeV.
For the $e\mu$ channel, we consider two types of events:
(i) events with  at least two jets ($e\mu$ 2-jets) and (ii)
events that contain just one detected jet ($e\mu$ 1-jet).

\item 
\label{sel:btag}
The $\ttbar$ final state contains two $b$-quark jets.
To improve separation between signal and background,
we apply a selection on the  
value of the MVA discriminant that assigns the $b$-quark hypothesis to
the two jets of largest $p_T$.
We use different cutoffs of the MVA discriminant variable, corresponding
 to $b$-jet efficiencies of 84\%  in $e\mu$ 2-jets, 80\% in $ee$, 
78\%  in $\mu\mu$, and 60\% in $e\mu$ 1-jet events, with background misidentification efficiencies, respectively, of 23\%, 12\%, 7\%, and 4\%.

\item
\label{sel:topo}
To improve signal purity, additional 
selection criteria are implemented based on global event properties of the final state.
In the $e\mu$ 1-jet events, we require $H_T>85$~GeV, 
where $H_T$ is the scalar sum of the transverse momenta of the leading lepton and the leading jet.
In the $e\mu$ 2-jets events, we require $H_T>108$~GeV, where $H_T$ is the scalar sum of the transverse momenta of the leading lepton and the two leading jets.
In the $ee$ final state, we require \metsig$>5$,
while in the $\mu\mu$ channel, we require  \met$>40$~GeV and \metsig$>2.5$.

\item
\label{sel:sameevts}
All leptons must have  $|\eta|<2$ and a difference in rapidity of $|\Delta\eta|<2.4$.
These criteria reduce the statistical uncertainty on the calculated parton-level asymmetries (see Sec.~\ref{sec:measurements}).
\end{enumerate}
The cut-off values of the selection criteria in items (\ref{sel:btag}) and (\ref{sel:topo}) 
are determined by minimizing the statistical uncertainty on the background-subtracted 
asymmetries (defined in Sec.~\ref{sec:measurements}).

To estimate the \ttbar\ signal efficiency and the background contamination,
we use MC simulation for all contributions except for the instrumental background, 
which is estimated from data.

In the $ee$ and $e\mu$ channels, we determine the contributions from events
with jets misidentified as electrons using the ``matrix method"~\cite{Abazov:2007kg}.
The loose sample of events ($n_{\text{loose}}$) is defined following
the same selection as used for the \ttbar\ candidate sample in items (\ref{sel:first})~--~(\ref{sel:topo}) above,
but ignoring the requirement on the electron MVA discriminant.
For the dielectron channel, we drop the MVA requirement on one of the electrons chosen randomly.

We measure the efficiency $\varepsilon_e$
that events with a true electron pass the requirement on the electron MVA discriminant  using \Z$\to ee$ data.
We measure the efficiency $f_e$  that events with a misidentified jet pass the electron MVA requirement
using $e\mu$ events chosen with selection criteria items (\ref{sel:first})~--~(\ref{sel:jets}),
but requiring leptons of the same electric charge.
For muons, we also apply a reversed isolation requirement:
$E^{\mu,\text{iso}}_T/p^{\mu}_T > 0.2$, $p^{\mu,\text{iso}}_T/p^{\mu}_T > 0.2$,
and $\met<15$~GeV, to minimize the contribution from $W$+jets events.

We extract the number of events with jets misidentified as electrons ($n_{f}$), and 
the number of events with true electrons ($n_e$), by solving the equations:
\begin{equation}
n_\text{{loose}} = n_{e}/\varepsilon_e + n_{f}/f_e ,
\end{equation}
\begin{equation}
n_{\text{tight}} = n_{e} + n_{f} ,
\end{equation}
where $n_{\text{tight}}$ is the number of events remaining after implementing the selections (\ref{sel:first})~--~(\ref{sel:topo}). 
The factors $f_e$ and $\varepsilon_e$ are measured separately for each jet multiplicity (0, 1, and 2 jets),
and separately for electron candidates in the CC and EC parts of the calorimeter.
Typical values of $\varepsilon_e$ are 0.7~--~0.8 in the CC and
0.65~--~0.75 in the EC. Values of $f_e$ are 0.005~--~0.010 in the CC, and
0.005~--~0.020 in the EC.

In the $e\mu$ and $\mu\mu$ channels, we determine the number of events 
with an isolated muon arising from decays of hadrons in jets 
relying on the same selection as for the $e\mu$ or $\mu\mu$ channels, but
requiring that both leptons have the same charge. 
In the $\mu\mu$ channel, this number of events is taken to be 
the number of same-sign events. 
In  the $e\mu$ channel, 
it is the number of events in the same-sign sample after subtracting the contribution from 
 events with jets misidentified as electrons.

The numbers of predicted background events, as well as the expected numbers of signal 
events, in the four channels are given in Table~\ref{tab:yield} and show high signal purity of the selected sample.

To complete the asymmetry measurement, we must determine not only the total number of 
events arising from instrumental background, but also 
their distributions in  $q\times\eta$ and $\Delta\eta$.
To determine these distributions for this background in the $ee$ and $e\mu$ channels, we use the loose
selection described above and implement a veto on events with one tight electron ($e\mu$ channel)
or two tight electrons ($ee$ channel). The residual contributions of the $Z$ boson and diboson
processes, as well as the expected contribution from the $\ttbar$ events, are subtracted.
In the $\mu\mu$ channel, we use the same sign events, where each of the muons is taken to have 
alternatively a negative and positive charge.
The resulting distributions are normalized to the number of previously estimated background events.

\begin{table*}
\renewcommand{\arraystretch}{1.8}
\caption{Numbers of total expected ($N_{\rm expected}$) and observed ($N_{\rm observed}$) events
from backgrounds and \ttbar\ signal assuming the SM cross section (7.45~pb for a top quark mass of $m_t=172.5$~GeV~\cite{PhysRevD.78.034003}).
Expected numbers of events are shown with their statistical uncertainties.
The uncertainty on the ratio of $N_{\rm observed}$/$N_{\rm expected}$
takes into account the statistical uncertainty on $N_{\rm observed}$ and $N_{\rm expected}$.
\label{tab:yield}}
\begin{tabular}[t]{lccccccc}
\hline\hline
& $Z\to \ell\ell$  & Dibosons& \parbox{2.2cm}{Multijet and $W$+jets}
& \parbox{2.2cm}{$t\bar{t}\to \ell\ell jj$} & \parbox{1.6cm}{$N_{\rm expected}$}
& \parbox{1.3cm}{$N_{\rm observed}$}& \parbox{1.3cm}{$\frac{N_{\rm observed}}{N_{\rm expected}}$} \\ \hline
$ee$ & $17.2^{+0.6}_{-0.6}$  & $2.4^{+0.1}_{-0.1}$  & $\phantom{0}4.7^{+0.4}_{-0.4}$  & $127.8^{-1.4}_{-1.4}$  & $152.1^{+1.6}_{-1.6}$  & 147 & $0.97\pm0.08$\\ 
$e\mu$ 2 jets & $13.7^{+0.5}_{-0.5}$ & $3.9^{+0.2}_{-0.2}$  & $16.3^{+4.0}_{-4.0}$  & $314.7^{+1.1}_{-1.1}$  & $348.6^{+4.2}_{-4.2}$  & 343 & $0.98\pm0.05$\\ 
$e\mu$ 1 jet & $\phantom{0}8.7^{+0.6}_{-0.6}$ & $3.4^{+0.2}_{-0.2}$  & $\phantom{0}2.9^{+1.7}_{-1.7}$  & $\phantom{0}61.7^{+0.5}_{-0.5}$  & $\phantom{0}76.7^{-1.9}_{-1.9}$  & \phantom{0}78 & $1.02\pm0.12$\\ 
$\mu\mu$ & $17.5^{+0.6}_{-0.6}$  & $1.9^{+0.1}_{-0.1}$ &  $\phantom{0}0.0^{+0.0}_{-0.0}$  & $\phantom{0}97.7^{+0.6}_{-0.6}$  & $117.1^{+0.8}_{-0.8}$  & 114 & $0.97\pm0.09$\\ \hline\hline
\end{tabular}

\end{table*}

\section{Method}
\label{sec:measurements}
\begin{figure}
\includegraphics[width=.40\textwidth]{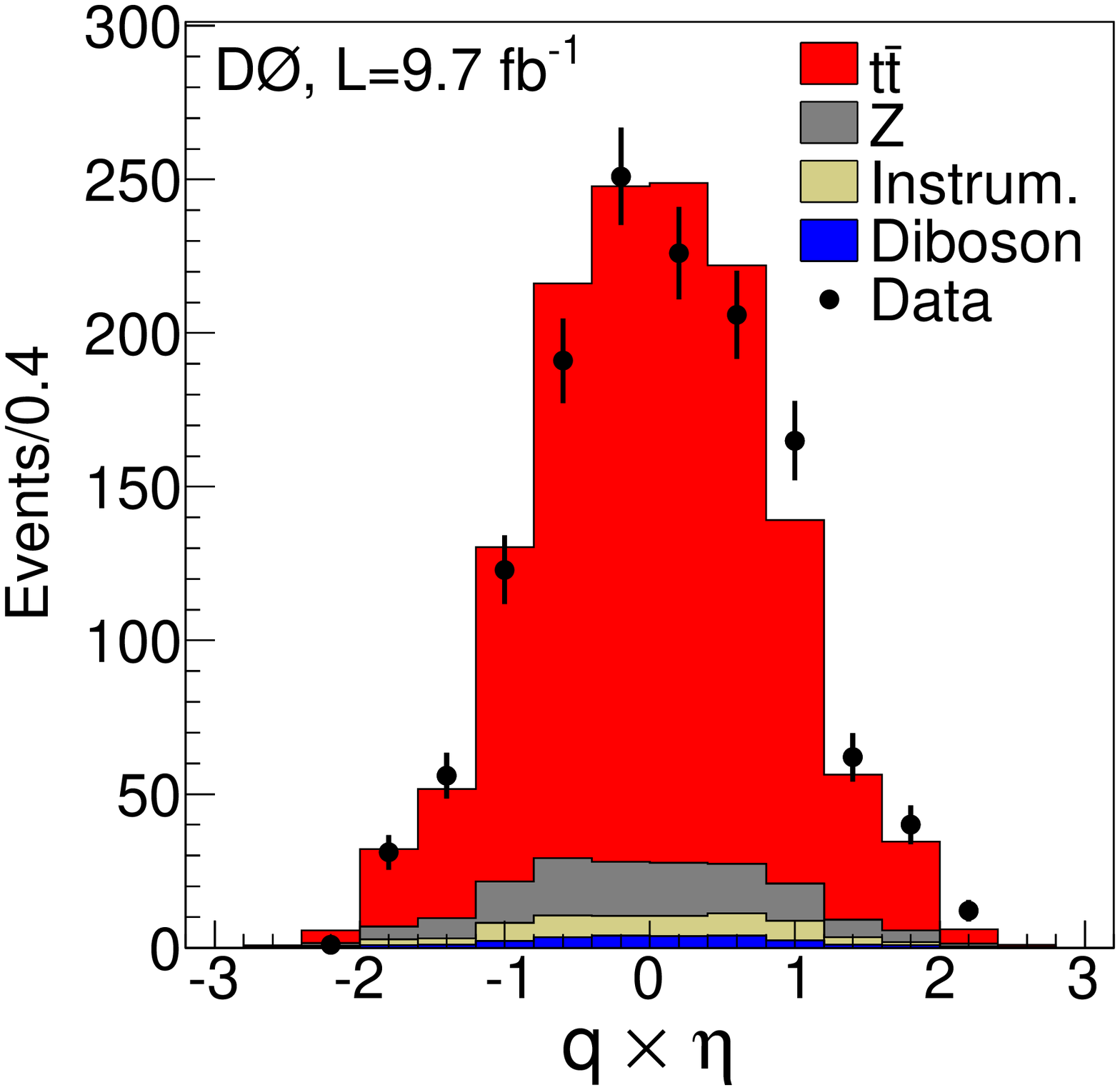}
\includegraphics[width=.40\textwidth]{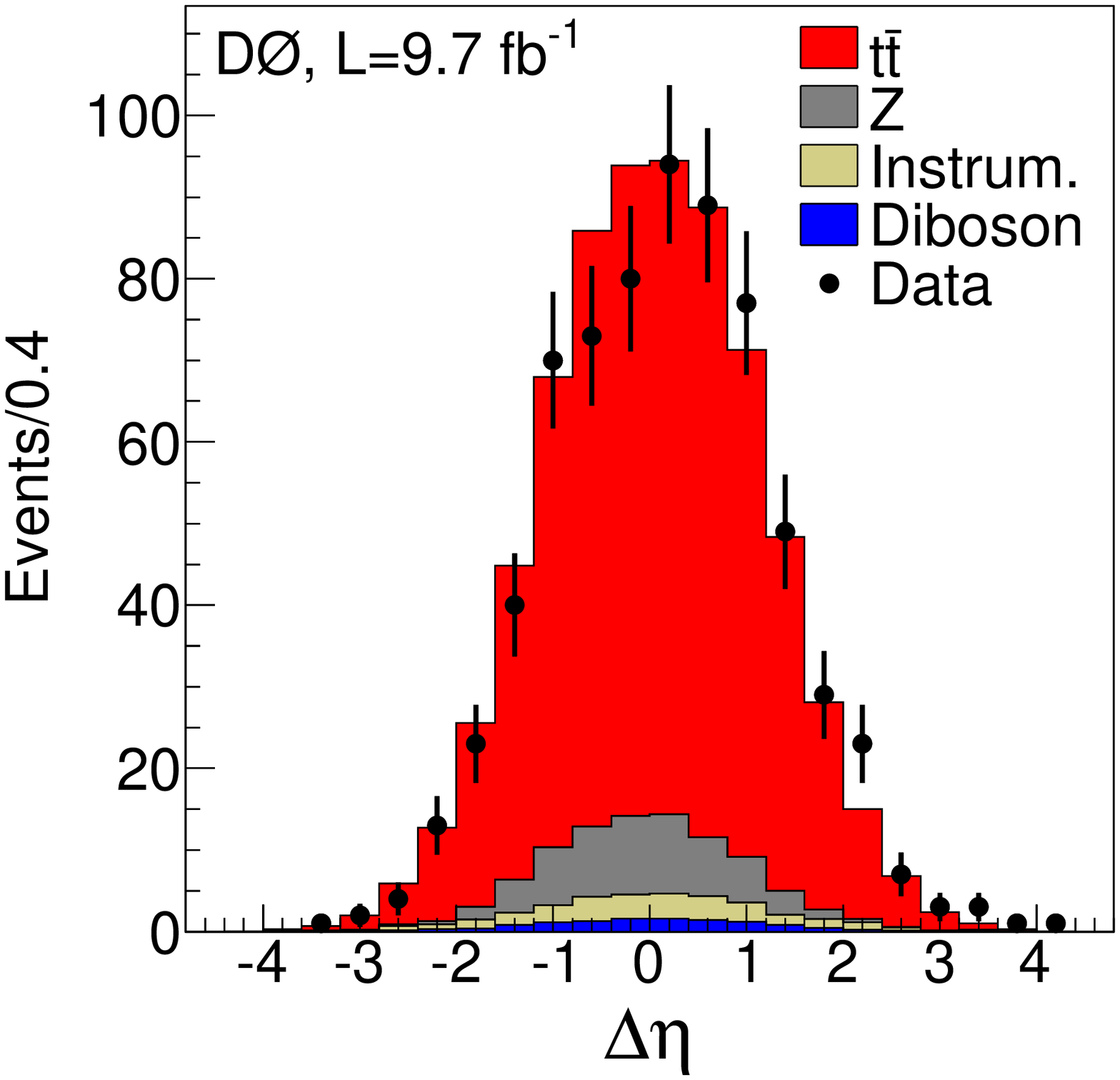}

\begin{picture}(00,00)

\put(-55,365){\text{\bf (a)}} 
\put(-55,168){\text{\bf (b)}} 

\end{picture} 

\caption{(color online) Distributions in (a) $q\times\eta$ and (b) $\Delta\eta = \eta_{\ell^+} -\eta_{\ell^-}$, for the 
sum of $ee$, $e\mu$ and $\mu\mu$ channels, along with predictions of the backgrounds and \ttbar\ signal.
The black points show data events and the error bars indicate the statistical uncertainty on the data.
\label{fig:rec_level}}
\end{figure}

\begin{figure}
\includegraphics[width=.40\textwidth]{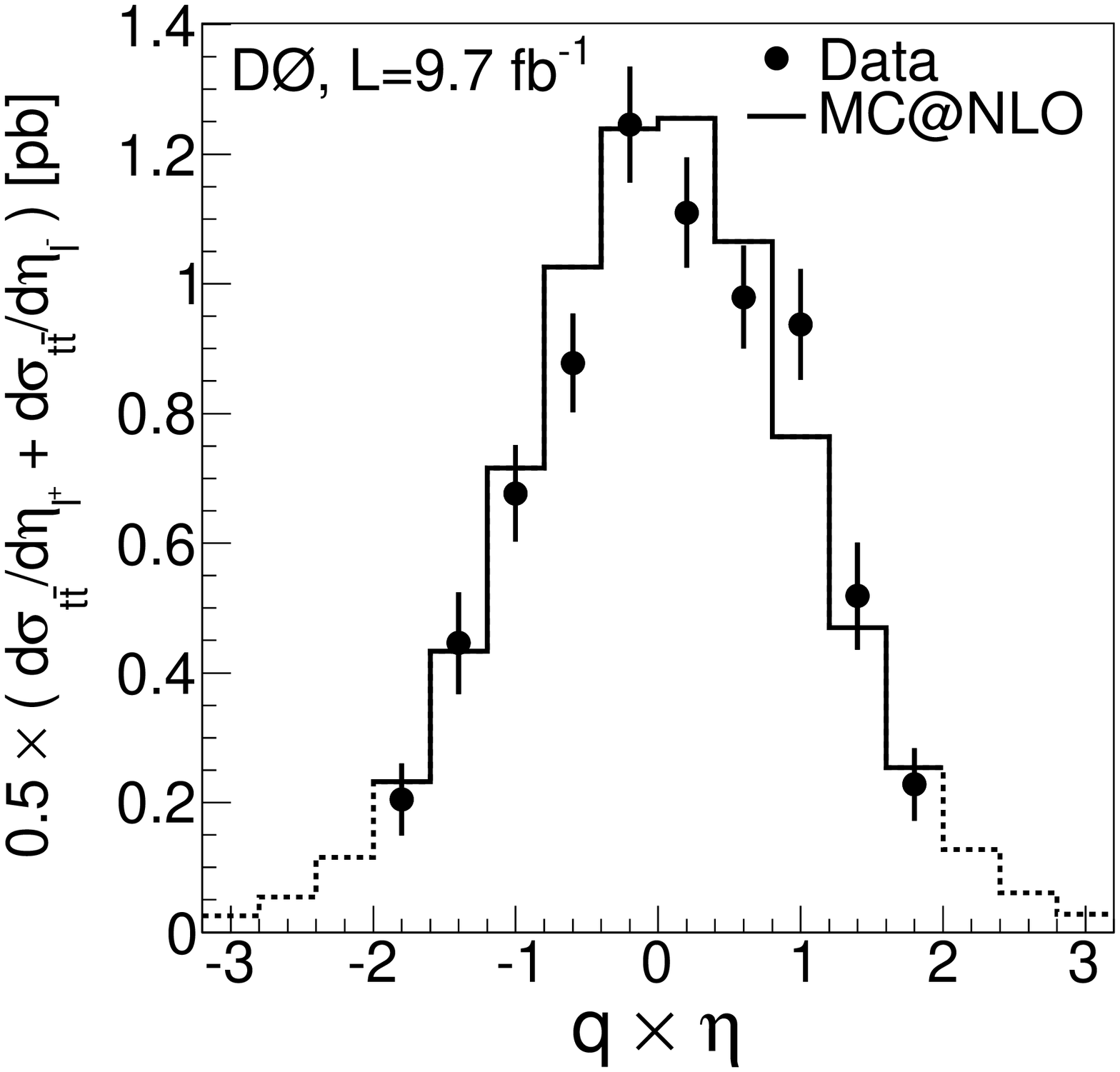}
\includegraphics[width=.40\textwidth]{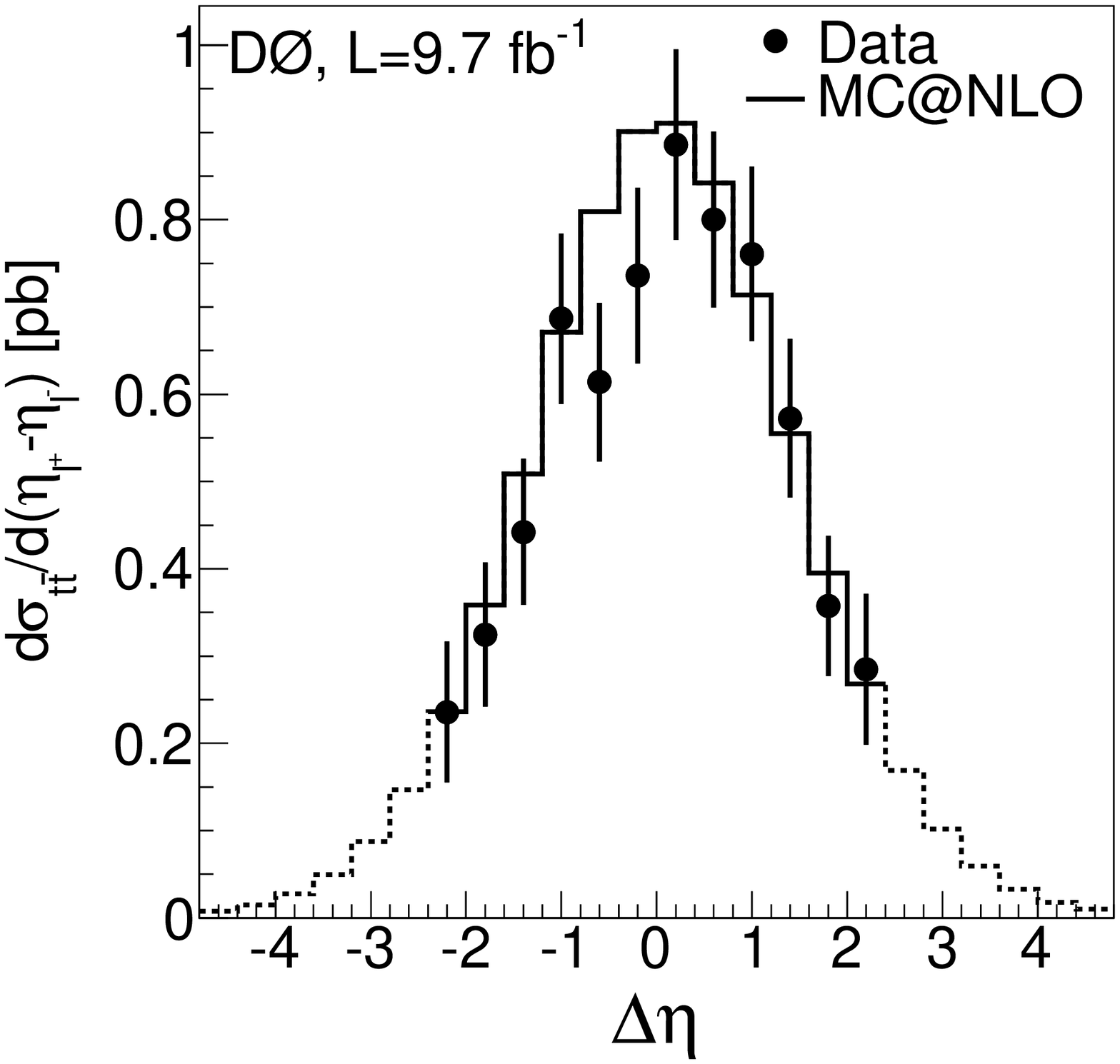}

\begin{picture}(00,00)

\put(-55,365){\text{\bf (a)}} 
\put(-55,168){\text{\bf (b)}} 

\end{picture}

\caption{Distributions in (a) $q\times\eta$ and (b) $\Delta\eta$,
for the combined $ee$, $e\mu$, and $\mu\mu$ channels after subtraction of background and correction for selection efficiency
within the acceptance. 
The error bars indicate the statistical uncertainty on data.
The dashed lines show the predictions from \mcatnlo\ outside the analysis acceptance.
\label{fig:corr_level}}
\end{figure}

Figure~\ref{fig:rec_level} presents the  $q\times\eta$ and $\Delta\eta$ 
distributions for dilepton events after applying all but item~(\ref{sel:sameevts}) of the selection criteria.
We compute \Alfb\ and \All\ in two steps.
First, within each of the four channels,  we perform a bin-by-bin subtraction of the estimated  background contributions to the data.
The lepton pseudorapidities are measured in \dzero\ with a resolution better than 1\% resulting in negligible migration effects.
We therefore apply a simple bin-by-bin correction, which suffices to account for the efficiency of reconstruction and selection requirements.
The correction function is determined using \ttbar\ \mcatnlo\ events at the parton level within the fiducial region
$|\eta|<2$, $|\Delta\eta|<2.4$ (here $\eta$ refers to the generated lepton pseudorapidity) 
and events after reconstruction and selection.
The asymmetries in the $q\times\eta$ and $\Delta\eta$ distributions after correction for selection efficiency 
are referred as
``corrected'' asymmetries.
Figure~\ref{fig:corr_level} shows the corrected distributions for data compared to the predictions from \mcatnlo.
The cross section in each bin is calculated as a weighted sum of the measurements in all channels, where only the
statistical uncertainty is taken into account.

In the second step, we extrapolate the corrected asymmetries to the full range of $\eta$
by multiplying the corrected asymmetries with the calculated extrapolation factor,
which is given by the ratio of
the generator level SM \ttbar\ asymmetries from \mcatnlo\, without selections to 
asymmetries within the fiducial region ($|\eta|<2$ and $|\Delta\eta|<2.4$).
We refer to these asymmetries as ``extrapolated'' asymmetries.
The exact values of the $|\eta|$ and $|\Delta\eta|$ requirements are chosen to optimize the expected statistical 
precision of the extrapolated asymmetries.

\section{Systematic Uncertainties}
\label{sec:systematics}

Systematic effects can affect the measured asymmetries in different ways: (i) they can change the
normalization or the differential dependence, i.e., ``shape'', of the background distributions, (ii) they can affect the
efficiency corrections and thereby modify the corrected and extrapolated asymmetries, 
and (iii) different MC generators or model assumptions can impact the extrapolation to all phase space. 
For item (iii), we verify that when axigluon MC samples (see Sec~\ref{sec:samples} and Table~\ref{tab:axigluons} for predicted asymmetries) are used instead of \mcatnlo\ 
to compute the extrapolation factor, we get consistent extrapolation factors.
This shows that the model assumed for the extrapolation does not significantly affect the 
extrapolated correction.

We first consider the following sources of the systematic uncertainty:
uncertainties on the efficiencies of electron and muon identification, uncertainties on trigger efficiencies,
and uncertainties on jet-related 
quantities. The latter  include contributions from the uncertainty in jet energy scale, jet energy resolution, jet identification 
efficiency, and $b$-quark jet tagging efficiency. All of these systematic uncertainties are propagated to the
background distributions and to the corrections for \ttbar\ signal efficiency; they are found to be small and
are grouped into the object identification (Object ID) category.

Next, we consider uncertainties specific to the background model. These include uncertainties 
on the asymmetries generated for $Z$ boson events (see Sec.~\ref{sec:samples}) and
on background normalization, which is typically $\approx 10$\%.
The background normalization uncertainty accounts for the uncertainties on the integrated luminosity~\cite{Andeen:2007zc}, object ID efficiency, 
$b$-tagging identification efficiency, and theoretical background cross sections.
We also calibrate our ability to reconstruct angular asymmetries by comparing asymmetries observed for $Z$ bosons in data with MC simulation. 
We use samples with requirements~(\ref{sel:first})--(\ref{sel:incl}) only
and ignore any jet selection in order to have a significant number of events and therefore a small
statistical uncertainty on the asymmetry ($\approx$0.13\% in data and $\approx$0.04\% in simulation).
We verify that  we can reproduce the asymmetries observed in data if we reweight the MC distributions using distributions 
obtained with \pythia~8~\cite{Sjostrand2008}.
This reweighting is based on the ratio of two-dimensional distributions in ($\eta_{\ell^+},\eta_{\ell^-}$) space for \alpgen\ and \pythia~8.
After requiring one or two jets, we observe a residual difference 
between the data and MC asymmetries in a sample dominated by background obtained by reversing the $b$-quark-tagging requirement~(\ref{sel:btag}).
We take this difference as a systematic uncertainty on the contribution from the $Z$ boson background.

The most significant contribution to the background-related uncertainty is from the uncertainty on instrumental background. 
We estimate this by changing the 
amount of instrumental background according to the uncertainty on its normalization. 
We also account for possible uncertainties in the distribution of instrumental background
by changing the number of events in each bin of the of this instrumental background distribution by $\pm$~1~SD of its statistical uncertainty.
The changes are applied in opposite directions for the positive $q\times\eta\ge 0$ or $\Delta\eta\ge 0$ and negative $q\times\eta<0$ or $\Delta\eta<0$
parts of the distributions in order to maximize the effect.

Another important uncertainty is related to the choice of parton showering and hadronization in \ttbar\ events.
This is evaluated by taking the difference between the asymmetries obtained  with efficiency corrections and extrapolation factors using
\mcatnlo+\herwig\ and \alpgen+\pythia. This estimation also includes the difference in the simulation of NLO effects between \mcatnlo\ and \alpgen\ generators.

Finally, we consider the limited statistics of the MC samples used to measure the efficiency correction.
These provide the smallest contributions to the systematic uncertainties on the extracted asymmetries.
All the above systematic uncertainties are listed in Table~\ref{tab:syst}.

As shown in the following section, the main uncertainty on the measured asymmetries is due to the limited size of the data sample.

\begin{table}
\caption{Systematic uncertainties for the corrected and the extrapolated asymmetries.
All values are given in \%.
\label{tab:syst}}
\begin{tabular}[t]{lcccc}
\hline 
\hline
& \multicolumn{2}{c}{Corrected} & \multicolumn{2}{c}{Extrapolated} \\
& \multicolumn{2}{c}{\Alfb~ \All} & \multicolumn{2}{c}{\Alfb~ \All}  \\
\hline
Source &  & & & \\
\hline
Object ID &  \multicolumn{2}{c}{0.54~ 0.50} & \multicolumn{2}{c}{0.59~ 0.60}\\ 
Background &  \multicolumn{2}{c}{0.66~ 0.74} & \multicolumn{2}{c}{0.72~ 0.88} \\ 
Hadronization & \multicolumn{2}{c}{0.52~ 0.62} & \multicolumn{2}{c}{0.62~ 0.92} \\ 
MC statistics & \multicolumn{2}{c}{0.19~ 0.23} & \multicolumn{2}{c}{0.23~ 0.37}\\ 
\hline
Total & \multicolumn{2}{c}{1.02~ 1.12} & \multicolumn{2}{c}{1.14~ 1.46} \\
\hline
\hline
\end{tabular}

\end{table}

\section{Results}
\label{sec:results}

We combine  the asymmetries measured in the $ee$, $e\mu$ 2 jets, $e\mu$ 1 jet, and $\mu\mu$ channels using the BLUE method~\cite{Valassi2003391,Lyons1988110},
assuming 100\% correlation among their systematic uncertainties.
Table~\ref{tab:results} summarizes the corrected and extrapolated asymmetries, as well as  
the prediction from a SM NLO calculation including QCD and electroweak (EW) corrections~\cite{Bernreuther:2012sx}.
The measured values are consistent with theoretical predictions based on the SM.

\begin{table}
\caption{The measured corrected and extrapolated asymmetries defined in Eqs.~(\ref{eq:al}) and  (\ref{eq:all})
combined for all channels separately and combined, compared to the predicted SM NLO asymmetries~\cite{Bernreuther:2012sx} for inclusive \ttbar\ production.
The measured extrapolated asymmetry should be compared with the SM NLO prediction.
The first uncertainty on the measured values corresponds to the statistical and the second to the systematic
contribution. All values are given in \%.
The uncertainty on the SM NLO predictions are due to renormalization and factorization scale variations.
\label{tab:results}}
\renewcommand{\arraystretch}{1.2}
\begin{tabular}{lccc}
\hline
\hline 
\Alfb  & Corrected & Extrapolated & Prediction \\
$ee$ & \phantom{$-$}6.8 $\pm$ \phantom{1}8.5 $\pm$ 1.3 & & \\
$e\mu$ 2 jets & \phantom{$-$}5.0 $\pm$ \phantom{1}4.6 $\pm$ 1.0 & & \\
$e\mu$ 1 jet & $-$0.1 $\pm$ 10.4 $\pm$ 2.5 & & \\
$\mu\mu$ & \phantom{$-$}0.8 $\pm$ \phantom{1}8.5 $\pm$ 1.4 & & \\
Combined & \phantom{$-$}4.1 $\pm$ \phantom{1}3.5 $\pm$ 1.0 & \phantom{0}4.4 $\pm$ 3.7 $\pm$ 1.1 & 3.8 $\pm$ 0.3 \\
\hline
\All  & Corrected & Extrapolated & Prediction \\
$ee$ & \phantom{-}16.4 $\pm$ 10.4 $\pm$ 1.6 & & \\
$e\mu$ 2 jets & \phantom{-}11.1 $\pm$ \phantom{1}6.3 $\pm$ 1.3 & & \\
$e\mu$ 1 jet & $-$2.1 $\pm$ 15.7 $\pm$ 3.4 & & \\
$\mu\mu$ & \phantom{-1}7.4 $\pm$ 11.7 $\pm$ 1.4 & & \\
Combined & \phantom{-}10.5 $\pm$ \phantom{1}4.7 $\pm$ 1.1 & 12.3 $\pm$ 5.4 $\pm$ 1.5 & 4.8 $\pm$ 0.4 \\
\hline
\hline
\end{tabular}
\end{table}

\begin{figure}
\includegraphics[width=.40\textwidth]{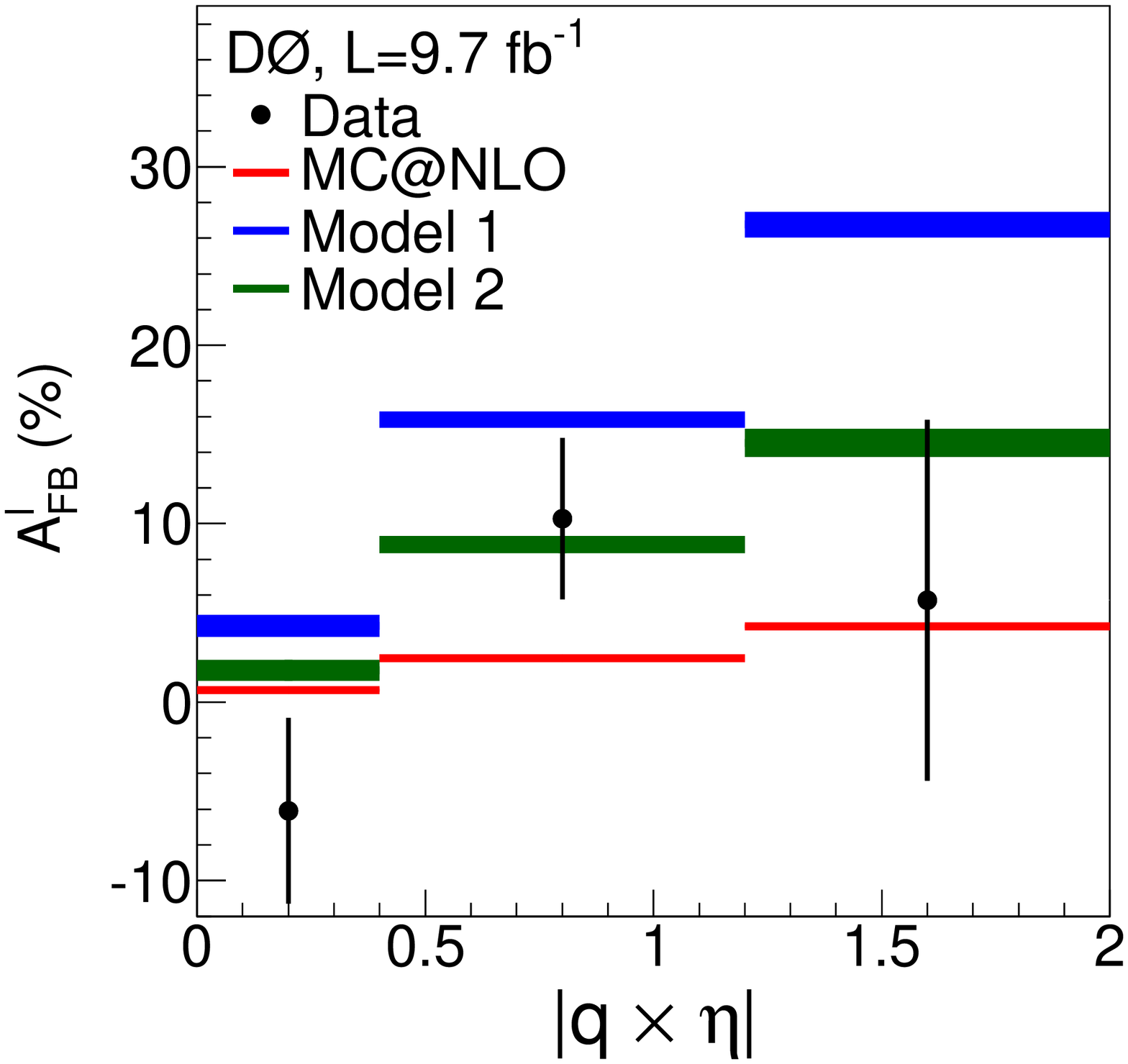}
\includegraphics[width=.40\textwidth]{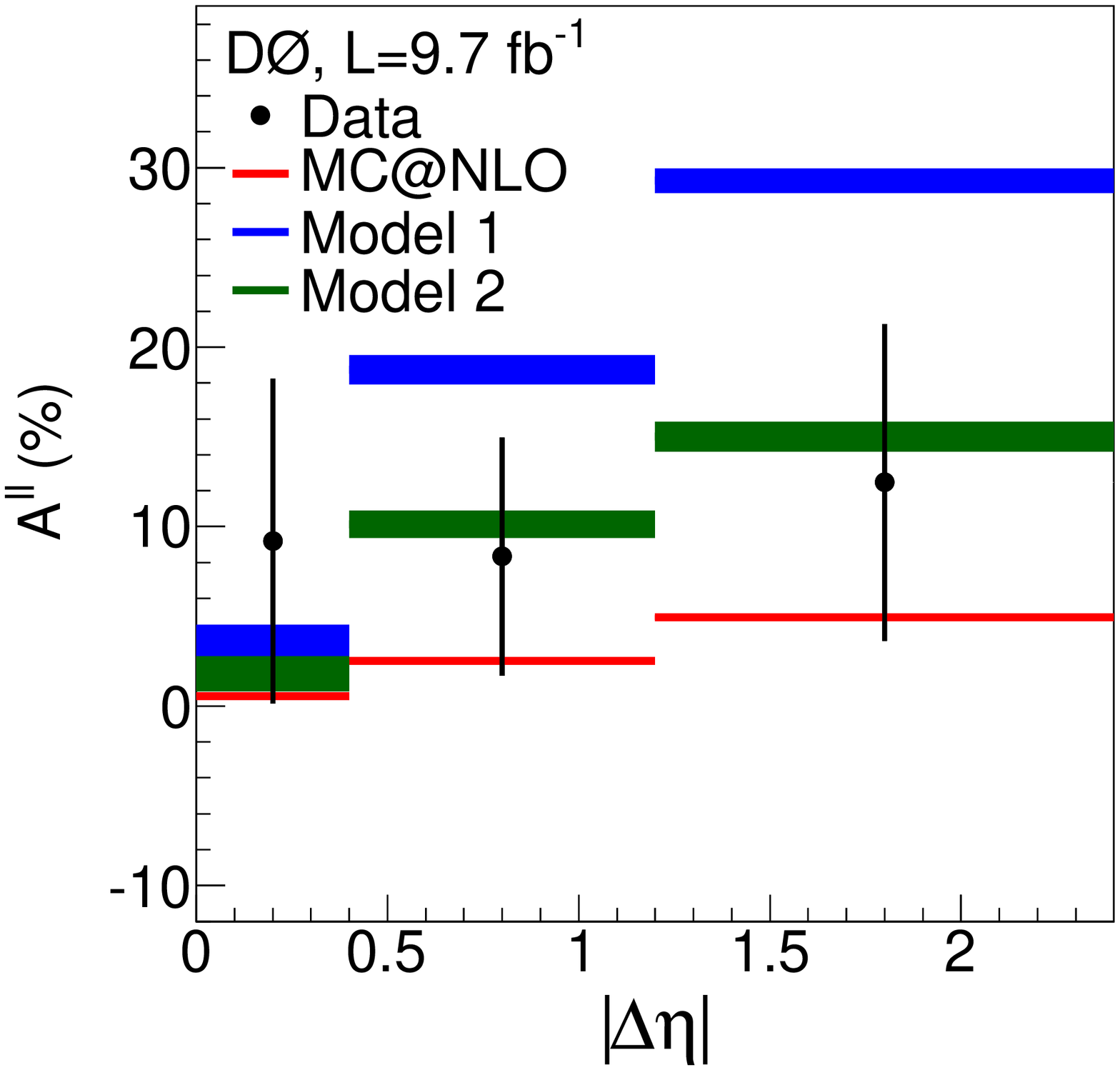}

\begin{picture}(00,00)

\put(70,380){\text{\bf (a)}} 
\put(70,183){\text{\bf (b)}} 

\end{picture}

\caption{(color online) Asymmetry distributions in (a) $|q\times\eta|$ and (b) $|\Delta\eta| = |\eta_{\ell^+} -\eta_{\ell^-}|$, for the
combined $ee$, $e\mu$, and $\mu\mu$ channels after background subtraction and after corrections for selection efficiency.
The error bars indicate statistical uncertainties on the data.
The data are compared with expectations from \mcatnlo\ and axigluon Model 1 and Model 2 as defined in the text.
\label{fig:asym_eta}}
\end{figure}

In addition, we study the dependence of the corrected asymmetries as a function of $q\times\eta$
and $\Delta\eta$ in Fig.~\ref{fig:asym_eta}, where 
we observe no significant dependence on these variables in the data and 
consistent with the \mcatnlo~\cite{Frixione:2002ik,Frixione:2008ym} predictions. Figure~\ref{fig:asym_eta} also shows the comparison with the two axigluon models described in 
Sec.~\ref{sec:samples}. 

\begin{figure}
\includegraphics[width=.40\textwidth]{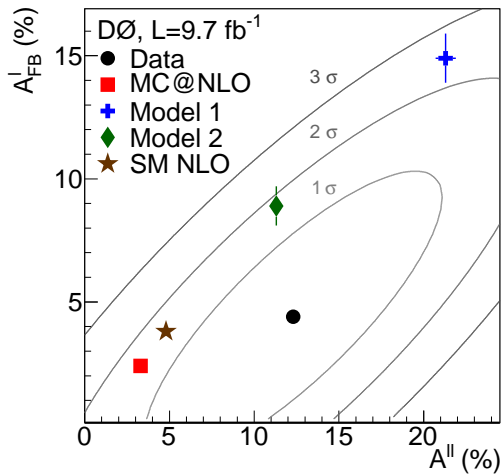}
\caption{Extrapolated \Alfb\ versus \All\ asymmetries in \ttbar\ data, the predictions from \mcatnlo,  axigluon models,
and from the latest SM NLO prediction~\cite{Bernreuther:2012sx}. The ellipses represent contours of total uncertainty 
at 1, 2, and 3 SD on the measured result. All values are given in \%.
Predicted asymmetries are shown with their statistical uncertainties.
\label{fig:al_vs_all}}
\end{figure}

To study the statistical correlation between \Alfb\ and \All, we
assume that positive and negative leptons have identical rapidity distributions, and we use the lepton $q\times \eta$ 
distribution in data (Fig.~\ref{fig:corr_level}) 
as the basis for generating an ensemble of $q\times \eta$ distributions. The residual 
reconstruction level differences between positive and negative leptons distributions  are made negligible 
by the regular flip of the solenoid and toroid polarities during the data taking.
The number of events in each bin is drawn from a Gaussian distribution with mean equal to the number of events in the bin 
of the initial distribution and width equal to the statistical uncertainty on the number of events.
The resulting distributions are used as probability density functions to generate  
pairs of rapidity values for positive and
negative leptons ($\eta_{l^+}, \eta_{l^-}$).
Since the value of $\eta$ for each lepton is generated independently, there is no direct correlation between them.
Repeating this  procedure many times, we form the $\Delta\eta=\eta_{\ell^+} - \eta_{\ell^-}$ distribution
and calculate both the \Alfb\ and \All\ asymmetries.
Using the (\Alfb, \All) pairs generated  in this way, we measure the correlation 
between the two asymmetries to be 0.82. 
We verify that the value of \All\ obtained with the same method but using the MC $q\times \eta$ event distribution as input
accurately reproduces the simulated asymmetry from \mcatnlo\ and axigluon models.
Using this correlation coefficient, we can compute the ratio of the two extrapolated asymmetries in data to be $R =  \Alfb/\All = 0.36 \pm 0.20$, consistent at the level
of 2 SD with the prediction of $0.79 \pm 0.10$.
The uncertainty on the theoretical ratio is estimated by adding in quadrature
the uncertainty on the theoretical expectations for \Alfb\ and \All\
and without taking into account the possible correlation between these two values.
This predicted ratio is found to be almost the same for the different tested models as can be seen in Fig.~\ref{fig:al_vs_all}.

The mean value of \All\ measured in this analysis differs from that in our previous measurement~\cite{Abazov:2012bfa},
but are compatible. The change in central value is due to 
changes in object identification and event selections (in particular, the use of $b$-quark jet identification) that improve the 
signal-to-background ratio and significantly reduce all systematic uncertainties related to background contributions,
which affects the central values of the results.

\section{Conclusion}
\label{sec:conclusion}
We have presented measurements of asymmetries in angular distributions of leptons 
produced in \ttbar\ dilepton final states.
Using the full Run~II Tevatron dataset recorded by the D0 detector, 
we measure the single lepton and dilepton asymmetries, 
corrected for reconstruction efficiency as:
\[\Alfb =  (4.1 \pm 3.5~(\text{stat}) \pm 1.0~(\text{syst}))\%\ ,  |\eta| < 2.0, |\Delta\eta| < 2.4, \] \text{and}
\[\All = (10.5 \pm 4.7~(\text{stat}) \pm 1.1~(\text{syst}))\%\ ,  |\eta| < 2.0, |\Delta\eta| < 2.4.\]
In addition, extrapolating these asymmetries for acceptance selections yields the inclusive \ttbar\ lepton asymmetries: 
\[\Alfb = (4.4 \pm 3.7~(\text{stat}) \pm 1.1~(\text{syst}))\%,\] \text{and}
\[\All = (12.3 \pm 5.4~(\text{stat}) \pm 1.5~(\text{syst}))\%.\]
These values are compatible with the SM NLO calculation that includes
QCD and EW corrections~\cite{Bernreuther:2012sx}.
We have studied the correlation between \Alfb\ and \All\ and 
computed the ratio of the two asymmetries, which also
shows agreement with calculations based on the standard model.

%

\section{acknowledgements}

We thank the staffs at Fermilab and collaborating institutions,
and acknowledge support from the
DOE and NSF (USA);
CEA and CNRS/IN2P3 (France);
MON, NRC KI and RFBR (Russia);
CNPq, FAPERJ, FAPESP and FUNDUNESP (Brazil);
DAE and DST (India);
Colciencias (Colombia);
CONACyT (Mexico);
NRF (Korea);
FOM (The Netherlands);
STFC and the Royal Society (United Kingdom);
MSMT and GACR (Czech Republic);
BMBF and DFG (Germany);
SFI (Ireland);
The Swedish Research Council (Sweden);
and
CAS and CNSF (China).
%

\section{Appendix: differential asymmetry tables}
\label{sec:tables}
Table~\ref{tab:corr} shows the \ttbar\ differential cross section in bins of $q\times\eta$ and $\Delta\eta$ as shown in Fig.~\ref{fig:corr_level}.
${1 \over 2} \times {({{d\sigma_{t\bar{t}}}\over{d\eta_{l^{+}}}} + {{d\sigma_{t\bar{t}}}\over{d\eta_{l^{-}} }})}$ represents the \ttbar\ differential cross section in $q\times\eta$
 and ${d\sigma_{t\bar{t}}}\over{d(\eta_{l^{+}}-\eta_{l^{-}})}$ the \ttbar\ differential cross section in $\Delta\eta$.
Table~\ref{tab:corr_diff} shows the values of the asymmetries in different angular regions as shown in Fig.~\ref{fig:asym_eta}.
\begin{table}[!h]
\caption{\ttbar\ cross section in each bin of Fig.~\ref{fig:corr_level}.
\label{tab:corr}}
\renewcommand{\arraystretch}{1.2}
\begin{tabular}[t]{ccc}
\hline
\hline
Bin & ${1 \over 2} \times {({{d\sigma_{t\bar{t}}}\over{d\eta_{l^{+}}}} + {{d\sigma_{t\bar{t}}}\over{d\eta_{l^{-}} }})}$ [pb] & ${d\sigma_{t\bar{t}}}\over{d(\eta_{l^{+}}-\eta_{l^{-}})}$ [pb] \\
\hline
$-2.4 , -2.0$ & 0.0 & 0.236 $\pm$ 0.081 \\
$-2.0 , -1.6$ & 0.205 $\pm$ 0.056 & 0.325 $\pm$ 0.082 \\
$-1.6 , -1.2$ & 0.446 $\pm$ 0.078 & 0.442 $\pm$ 0.084 \\
$-1.2 , -0.8$ & 0.677 $\pm$ 0.075 & 0.686 $\pm$ 0.097 \\
$-0.8 , -0.4$ & 0.878 $\pm$ 0.076 & 0.614 $\pm$ 0.091 \\
$-0.4 , \phantom{-}0.0$ & 1.245 $\pm$ 0.089 & 0.736 $\pm$ 0.101 \\
$\phantom{-}0.0 , \phantom{-}0.4$ & 1.110 $\pm$ 0.085 & 0.886 $\pm$ 0.109 \\
$\phantom{-}0.4 , \phantom{-}0.8$ & 0.979 $\pm$ 0.079 & 0.800 $\pm$ 0.101 \\
$\phantom{-}0.8 , \phantom{-}1.2$ & 0.937 $\pm$ 0.085 & 0.761 $\pm$ 0.100 \\
$\phantom{-}1.2 , \phantom{-}1.6$ & 0.518 $\pm$ 0.082 & 0.572 $\pm$ 0.091 \\
$\phantom{-}1.6 , \phantom{-}2.0$ & 0.228 $\pm$ 0.056 & 0.357 $\pm$ 0.081 \\
$\phantom{-}2.0 , \phantom{-}2.4$ & 0.0 & 0.285 $\pm$ 0.086 \\
\hline\hline
\end{tabular}

\end{table}
\begin{table}[!h]
\caption{Value of the asymmetries in different bins of the
distributions of Fig.~\ref{fig:asym_eta}.
\label{tab:corr_diff}}
\renewcommand{\arraystretch}{1.2}
\begin{tabular}[t]{cc}
\hline
\hline
$|q\times\eta|$ bin & \Alfb\ \\
\hline
$0.0 , 0.4$ & $-0.061$ $\pm$ 0.052 \\
$0.4 , 1.2$ & \phantom{-1}0.103 $\pm$ 0.045 \\
$1.2 , 2.0$ & \phantom{-1}0.057 $\pm$ 0.101 \\ 
\hline\hline
\end{tabular}
\begin{tabular}[t]{cc}
\hline
\hline
$|\Delta\eta|$ bin & \All\ \\
\hline
$0.0 , 0.4$ & 0.092 $\pm$ 0.091 \\
$0.4 , 1.2$ & 0.083 $\pm$ 0.066 \\
$1.2 , 2.4$ & 0.125 $\pm$ 0.088 \\
 \hline\hline
 \end{tabular}

\end{table}

\bibliography{references}

\begin{thebibliography}{65}
\expandafter\ifx\csname natexlab\endcsname\relax\def\natexlab#1{#1}\fi
\expandafter\ifx\csname bibnamefont\endcsname\relax
  \def\bibnamefont#1{#1}\fi
\expandafter\ifx\csname bibfnamefont\endcsname\relax
  \def\bibfnamefont#1{#1}\fi
\expandafter\ifx\csname citenamefont\endcsname\relax
  \def\citenamefont#1{#1}\fi
\expandafter\ifx\csname url\endcsname\relax
  \def\url#1{\texttt{#1}}\fi
\expandafter\ifx\csname urlprefix\endcsname\relax\def\urlprefix{URL }\fi
\providecommand{\bibinfo}[2]{#2}
\providecommand{\eprint}[2][]{\url{#2}}

\bibitem[{\citenamefont{Abe et~al.}(1995)}]{Abe:1995hr}
\bibinfo{author}{\bibfnamefont{F.}~\bibnamefont{Abe}}
  \bibnamefont{{\it{et~al.}}} (\bibinfo{collaboration}{CDF Collaboration}),
  \bibinfo{journal}{{Phys. Rev. Lett.}} \textbf{\bibinfo{volume}{74}},
  \bibinfo{pages}{2626} (\bibinfo{year}{1995}).

\bibitem[{\citenamefont{Abachi et~al.}(1995)}]{Abachi:1995iq}
\bibinfo{author}{\bibfnamefont{S.}~\bibnamefont{Abachi}}
  \bibnamefont{{\it{et~al.}}} (\bibinfo{collaboration}{D0 Collaboration}),
  \bibinfo{journal}{Phys. Rev. Lett.} \textbf{\bibinfo{volume}{74}},
  \bibinfo{pages}{2632} (\bibinfo{year}{1995}).

\bibitem[{\citenamefont{Bernreuther and Si}(2010)}]{Bernreuther:2010ny}
\bibinfo{author}{\bibfnamefont{W.}~\bibnamefont{Bernreuther}} \bibnamefont{and}
  \bibinfo{author}{\bibfnamefont{Z.-G.} \bibnamefont{Si}},
  \bibinfo{journal}{Nucl. Phys.} \textbf{\bibinfo{volume}{B837}},
  \bibinfo{pages}{90} (\bibinfo{year}{2010}).

\bibitem[{\citenamefont{Hollik and Pagani}(2011)}]{Hollik:2011ps}
\bibinfo{author}{\bibfnamefont{W.}~\bibnamefont{Hollik}} \bibnamefont{and}
  \bibinfo{author}{\bibfnamefont{D.}~\bibnamefont{Pagani}},
  \bibinfo{journal}{Phys. Rev. D} \textbf{\bibinfo{volume}{84}},
  \bibinfo{pages}{093003} (\bibinfo{year}{2011}).

\bibitem[{\citenamefont{Ahrens et~al.}(2011)\citenamefont{Ahrens, Ferroglia,
  Neubert, Pecjak, and Yang}}]{Ahrens:2011uf}
\bibinfo{author}{\bibfnamefont{V.}~\bibnamefont{Ahrens}},
  \bibinfo{author}{\bibfnamefont{A.}~\bibnamefont{Ferroglia}},
  \bibinfo{author}{\bibfnamefont{M.}~\bibnamefont{Neubert}},
  \bibinfo{author}{\bibfnamefont{B.~D.} \bibnamefont{Pecjak}},
  \bibnamefont{and} \bibinfo{author}{\bibfnamefont{L.~L.} \bibnamefont{Yang}},
  \bibinfo{journal}{Phys. Rev. D} \textbf{\bibinfo{volume}{84}},
  \bibinfo{pages}{074004} (\bibinfo{year}{2011}).

\bibitem[{\citenamefont{Kuhn and Rodrigo}(2012)}]{Kuhn:2011ri}
\bibinfo{author}{\bibfnamefont{J.~H.} \bibnamefont{Kuhn}} \bibnamefont{and}
  \bibinfo{author}{\bibfnamefont{G.}~\bibnamefont{Rodrigo}},
  \bibinfo{journal}{High Energy Phys.} \textbf{\bibinfo{volume}{1201}}
  (\bibinfo{year}{2012}), \bibinfo{note}{063}.

\bibitem[{\citenamefont{Ferrario and Rodrigo}(2008)}]{Ferrario:2008wm}
\bibinfo{author}{\bibfnamefont{P.}~\bibnamefont{Ferrario}} \bibnamefont{and}
  \bibinfo{author}{\bibfnamefont{G.}~\bibnamefont{Rodrigo}},
  \bibinfo{journal}{Phys. Rev. D} \textbf{\bibinfo{volume}{78}},
  \bibinfo{pages}{094018} (\bibinfo{year}{2008}).

\bibitem[{\citenamefont{Ferrario and Rodrigo}(2009)}]{Ferrario:2009bz}
\bibinfo{author}{\bibfnamefont{P.}~\bibnamefont{Ferrario}} \bibnamefont{and}
  \bibinfo{author}{\bibfnamefont{G.}~\bibnamefont{Rodrigo}},
  \bibinfo{journal}{Phys. Rev. D} \textbf{\bibinfo{volume}{80}},
  \bibinfo{pages}{051701} (\bibinfo{year}{2009}).

\bibitem[{\citenamefont{Antunano et~al.}(2008)\citenamefont{Antunano, Kuhn, and
  Rodrigo}}]{Antunano:2007da}
\bibinfo{author}{\bibfnamefont{O.}~\bibnamefont{Antunano}},
  \bibinfo{author}{\bibfnamefont{J.~H.} \bibnamefont{Kuhn}}, \bibnamefont{and}
  \bibinfo{author}{\bibfnamefont{G.}~\bibnamefont{Rodrigo}},
  \bibinfo{journal}{Phys. Rev. D} \textbf{\bibinfo{volume}{77}},
  \bibinfo{pages}{014003} (\bibinfo{year}{2008}).

\bibitem[{\citenamefont{Frampton et~al.}(2010)\citenamefont{Frampton, Shu, and
  Wang}}]{Frampton:2009rk}
\bibinfo{author}{\bibfnamefont{P.~H.} \bibnamefont{Frampton}},
  \bibinfo{author}{\bibfnamefont{J.}~\bibnamefont{Shu}}, \bibnamefont{and}
  \bibinfo{author}{\bibfnamefont{K.}~\bibnamefont{Wang}},
  \bibinfo{journal}{Phys. Lett. B} \textbf{\bibinfo{volume}{683}},
  \bibinfo{pages}{294} (\bibinfo{year}{2010}).

\bibitem[{\citenamefont{Jung et~al.}(2010{\natexlab{a}})\citenamefont{Jung, Ko,
  Lee, and Nam}}]{Jung:2009pi}
\bibinfo{author}{\bibfnamefont{D.-W.} \bibnamefont{Jung}},
  \bibinfo{author}{\bibfnamefont{P.}~\bibnamefont{Ko}},
  \bibinfo{author}{\bibfnamefont{J.~S.} \bibnamefont{Lee}}, \bibnamefont{and}
  \bibinfo{author}{\bibfnamefont{S.-H.} \bibnamefont{Nam}},
  \bibinfo{journal}{Phys. Lett. B} \textbf{\bibinfo{volume}{691}},
  \bibinfo{pages}{238} (\bibinfo{year}{2010}{\natexlab{a}}).

\bibitem[{\citenamefont{Arhrib et~al.}(2010)\citenamefont{Arhrib, Benbrik, and
  Chen}}]{Arhrib:2009hu}
\bibinfo{author}{\bibfnamefont{A.}~\bibnamefont{Arhrib}},
  \bibinfo{author}{\bibfnamefont{R.}~\bibnamefont{Benbrik}}, \bibnamefont{and}
  \bibinfo{author}{\bibfnamefont{C.-H.} \bibnamefont{Chen}},
  \bibinfo{journal}{Phys. Rev. D} \textbf{\bibinfo{volume}{82}},
  \bibinfo{pages}{034034} (\bibinfo{year}{2010}).

\bibitem[{\citenamefont{Djouadi et~al.}(2010)\citenamefont{Djouadi, Moreau,
  Richard, and Singh}}]{Djouadi:2009nb}
\bibinfo{author}{\bibfnamefont{A.}~\bibnamefont{Djouadi}},
  \bibinfo{author}{\bibfnamefont{G.}~\bibnamefont{Moreau}},
  \bibinfo{author}{\bibfnamefont{F.}~\bibnamefont{Richard}}, \bibnamefont{and}
  \bibinfo{author}{\bibfnamefont{R.~K.} \bibnamefont{Singh}},
  \bibinfo{journal}{Phys. Rev. D} \textbf{\bibinfo{volume}{82}},
  \bibinfo{pages}{071702} (\bibinfo{year}{2010}).

\bibitem[{\citenamefont{Alvarez et~al.}(2011)\citenamefont{Alvarez, Da~Rold,
  Vietto, and Szynkman}}]{Alvarez:2011hi}
\bibinfo{author}{\bibfnamefont{E.}~\bibnamefont{Alvarez}},
  \bibinfo{author}{\bibfnamefont{L.}~\bibnamefont{Da~Rold}},
  \bibinfo{author}{\bibfnamefont{J.~I.~S.} \bibnamefont{Vietto}},
  \bibnamefont{and} \bibinfo{author}{\bibfnamefont{A.}~\bibnamefont{Szynkman}},
  \bibinfo{journal}{J. High Energy Phys.} \textbf{\bibinfo{volume}{09}}
  (\bibinfo{year}{2011}), \bibinfo{note}{007}.

\bibitem[{\citenamefont{Chen et~al.}(2011)\citenamefont{Chen, Cvetic, and
  Kim}}]{Chen:2010hm}
\bibinfo{author}{\bibfnamefont{C.-H.} \bibnamefont{Chen}},
  \bibinfo{author}{\bibfnamefont{G.}~\bibnamefont{Cvetic}}, \bibnamefont{and}
  \bibinfo{author}{\bibfnamefont{C.}~\bibnamefont{Kim}},
  \bibinfo{journal}{Phys. Lett. B} \textbf{\bibinfo{volume}{694}},
  \bibinfo{pages}{393} (\bibinfo{year}{2011}).

\bibitem[{\citenamefont{Jung et~al.}(2011)\citenamefont{Jung, Ko, and
  Lee}}]{Jung:2010yn}
\bibinfo{author}{\bibfnamefont{D.-W.} \bibnamefont{Jung}},
  \bibinfo{author}{\bibfnamefont{P.}~\bibnamefont{Ko}}, \bibnamefont{and}
  \bibinfo{author}{\bibfnamefont{J.~S.} \bibnamefont{Lee}},
  \bibinfo{journal}{Phys. Lett. B} \textbf{\bibinfo{volume}{701}},
  \bibinfo{pages}{248} (\bibinfo{year}{2011}).

\bibitem[{\citenamefont{Aguilar-Saavedra and
  Perez-Victoria}(2011{\natexlab{a}})}]{AguilarSaavedra:2011ci}
\bibinfo{author}{\bibfnamefont{J.}~\bibnamefont{Aguilar-Saavedra}}
  \bibnamefont{and}
  \bibinfo{author}{\bibfnamefont{M.}~\bibnamefont{Perez-Victoria}},
  \bibinfo{journal}{Phys. Lett. B} \textbf{\bibinfo{volume}{705}},
  \bibinfo{pages}{228} (\bibinfo{year}{2011}{\natexlab{a}}).

\bibitem[{\citenamefont{Marques~Tavares and Schmaltz}(2011)}]{Tavares:2011zg}
\bibinfo{author}{\bibfnamefont{G.}~\bibnamefont{Marques~Tavares}}
  \bibnamefont{and} \bibinfo{author}{\bibfnamefont{M.}~\bibnamefont{Schmaltz}},
  \bibinfo{journal}{Phys. Rev. D} \textbf{\bibinfo{volume}{84}},
  \bibinfo{pages}{054008} (\bibinfo{year}{2011}).

\bibitem[{\citenamefont{Barcelo et~al.}(2012)\citenamefont{Barcelo, Carmona,
  Masip, and Santiago}}]{Barcelo:2011vk}
\bibinfo{author}{\bibfnamefont{R.}~\bibnamefont{Barcelo}},
  \bibinfo{author}{\bibfnamefont{A.}~\bibnamefont{Carmona}},
  \bibinfo{author}{\bibfnamefont{M.}~\bibnamefont{Masip}}, \bibnamefont{and}
  \bibinfo{author}{\bibfnamefont{J.}~\bibnamefont{Santiago}},
  \bibinfo{journal}{Phys Lett. B} \textbf{\bibinfo{volume}{707}},
  \bibinfo{pages}{88} (\bibinfo{year}{2012}).

\bibitem[{\citenamefont{Cheung et~al.}(2009)\citenamefont{Cheung, Keung, and
  Yuan}}]{Cheung:2009ch}
\bibinfo{author}{\bibfnamefont{K.}~\bibnamefont{Cheung}},
  \bibinfo{author}{\bibfnamefont{W.-Y.} \bibnamefont{Keung}}, \bibnamefont{and}
  \bibinfo{author}{\bibfnamefont{T.-C.} \bibnamefont{Yuan}},
  \bibinfo{journal}{Phys. Lett. B} \textbf{\bibinfo{volume}{682}},
  \bibinfo{pages}{287} (\bibinfo{year}{2009}).

\bibitem[{\citenamefont{Xiao et~al.}(2010)\citenamefont{Xiao, Wang, and
  Zhu}}]{Xiao:2010hm}
\bibinfo{author}{\bibfnamefont{B.}~\bibnamefont{Xiao}},
  \bibinfo{author}{\bibfnamefont{Y.-k.} \bibnamefont{Wang}}, \bibnamefont{and}
  \bibinfo{author}{\bibfnamefont{S.-h.} \bibnamefont{Zhu}},
  \bibinfo{journal}{Phys. Rev. D} \textbf{\bibinfo{volume}{82}},
  \bibinfo{pages}{034026} (\bibinfo{year}{2010}).

\bibitem[{\citenamefont{Cao et~al.}(2010{\natexlab{a}})\citenamefont{Cao,
  McKeen, Rosner, Shaughnessy, and Wagner}}]{Cao:2010zb}
\bibinfo{author}{\bibfnamefont{Q.-H.} \bibnamefont{Cao}},
  \bibinfo{author}{\bibfnamefont{D.}~\bibnamefont{McKeen}},
  \bibinfo{author}{\bibfnamefont{J.~L.} \bibnamefont{Rosner}},
  \bibinfo{author}{\bibfnamefont{G.}~\bibnamefont{Shaughnessy}},
  \bibnamefont{and} \bibinfo{author}{\bibfnamefont{C.~E.}
  \bibnamefont{Wagner}}, \bibinfo{journal}{Phys. Rev. D}
  \textbf{\bibinfo{volume}{81}}, \bibinfo{pages}{114004}
  (\bibinfo{year}{2010}{\natexlab{a}}).

\bibitem[{\citenamefont{Jung et~al.}(2010{\natexlab{b}})\citenamefont{Jung,
  Murayama, Pierce, and Wells}}]{Jung:2009jz}
\bibinfo{author}{\bibfnamefont{S.}~\bibnamefont{Jung}},
  \bibinfo{author}{\bibfnamefont{H.}~\bibnamefont{Murayama}},
  \bibinfo{author}{\bibfnamefont{A.}~\bibnamefont{Pierce}}, \bibnamefont{and}
  \bibinfo{author}{\bibfnamefont{J.~D.} \bibnamefont{Wells}},
  \bibinfo{journal}{Phys. Rev. D} \textbf{\bibinfo{volume}{81}},
  \bibinfo{pages}{015004} (\bibinfo{year}{2010}{\natexlab{b}}).

\bibitem[{\citenamefont{Cao et~al.}(2010{\natexlab{b}})\citenamefont{Cao, Heng,
  Wu, and Yang}}]{Cao:2009uz}
\bibinfo{author}{\bibfnamefont{J.}~\bibnamefont{Cao}},
  \bibinfo{author}{\bibfnamefont{Z.}~\bibnamefont{Heng}},
  \bibinfo{author}{\bibfnamefont{L.}~\bibnamefont{Wu}}, \bibnamefont{and}
  \bibinfo{author}{\bibfnamefont{J.~M.} \bibnamefont{Yang}},
  \bibinfo{journal}{Phys. Rev. D} \textbf{\bibinfo{volume}{81}},
  \bibinfo{pages}{014016} (\bibinfo{year}{2010}{\natexlab{b}}).

\bibitem[{\citenamefont{Barger et~al.}(2010)\citenamefont{Barger, Keung, and
  Yu}}]{Barger:2010mw}
\bibinfo{author}{\bibfnamefont{V.}~\bibnamefont{Barger}},
  \bibinfo{author}{\bibfnamefont{W.-Y.} \bibnamefont{Keung}}, \bibnamefont{and}
  \bibinfo{author}{\bibfnamefont{C.-T.} \bibnamefont{Yu}},
  \bibinfo{journal}{Phys. Rev. D} \textbf{\bibinfo{volume}{81}},
  \bibinfo{pages}{113009} (\bibinfo{year}{2010}).

\bibitem[{\citenamefont{Bauer et~al.}(2010)\citenamefont{Bauer, Goertz, Haisch,
  Pfoh, and Westhoff}}]{Bauer:2010iq}
\bibinfo{author}{\bibfnamefont{M.}~\bibnamefont{Bauer}},
  \bibinfo{author}{\bibfnamefont{F.}~\bibnamefont{Goertz}},
  \bibinfo{author}{\bibfnamefont{U.}~\bibnamefont{Haisch}},
  \bibinfo{author}{\bibfnamefont{T.}~\bibnamefont{Pfoh}}, \bibnamefont{and}
  \bibinfo{author}{\bibfnamefont{S.}~\bibnamefont{Westhoff}},
  \bibinfo{journal}{J. High Energy Phys.} \textbf{\bibinfo{volume}{11}}
  (\bibinfo{year}{2010}), \bibinfo{note}{039}.

\bibitem[{\citenamefont{Chivukula et~al.}(2010)\citenamefont{Chivukula,
  Simmons, and Yuan}}]{Chivukula:2010fk}
\bibinfo{author}{\bibfnamefont{R.~S.} \bibnamefont{Chivukula}},
  \bibinfo{author}{\bibfnamefont{E.~H.} \bibnamefont{Simmons}},
  \bibnamefont{and} \bibinfo{author}{\bibfnamefont{C.-P.} \bibnamefont{Yuan}},
  \bibinfo{journal}{Phys. Rev. D} \textbf{\bibinfo{volume}{82}},
  \bibinfo{pages}{094009} (\bibinfo{year}{2010}).

\bibitem[{\citenamefont{Dorsner et~al.}(2009)\citenamefont{Dorsner, Fajfer,
  Kamenik, and Kosnik}}]{Dorsner:2009mq}
\bibinfo{author}{\bibfnamefont{I.}~\bibnamefont{Dorsner}},
  \bibinfo{author}{\bibfnamefont{S.}~\bibnamefont{Fajfer}},
  \bibinfo{author}{\bibfnamefont{J.~F.} \bibnamefont{Kamenik}},
  \bibnamefont{and} \bibinfo{author}{\bibfnamefont{N.}~\bibnamefont{Kosnik}},
  \bibinfo{journal}{Phys. Rev. D} \textbf{\bibinfo{volume}{81}},
  \bibinfo{pages}{055009} (\bibinfo{year}{2009}).

\bibitem[{\citenamefont{Shu et~al.}(2010)\citenamefont{Shu, Tait, and
  Wang}}]{Shu:2009xf}
\bibinfo{author}{\bibfnamefont{J.}~\bibnamefont{Shu}},
  \bibinfo{author}{\bibfnamefont{T.~M.} \bibnamefont{Tait}}, \bibnamefont{and}
  \bibinfo{author}{\bibfnamefont{K.}~\bibnamefont{Wang}},
  \bibinfo{journal}{Phys. Rev. D} \textbf{\bibinfo{volume}{81}},
  \bibinfo{pages}{034012} (\bibinfo{year}{2010}).

\bibitem[{\citenamefont{Aguilar-Saavedra and
  Perez-Victoria}(2011{\natexlab{b}})}]{AguilarSaavedra:2011ug}
\bibinfo{author}{\bibfnamefont{J.}~\bibnamefont{Aguilar-Saavedra}}
  \bibnamefont{and}
  \bibinfo{author}{\bibfnamefont{M.}~\bibnamefont{Perez-Victoria}},
  \bibinfo{journal}{J. High Energy Phys.} \textbf{\bibinfo{volume}{09}}
  (\bibinfo{year}{2011}{\natexlab{b}}), \bibinfo{note}{097}.

\bibitem[{\citenamefont{Aaltonen et~al.}(2013{\natexlab{a}})}]{Aaltonen:2012it}
\bibinfo{author}{\bibfnamefont{T.}~\bibnamefont{Aaltonen}}
  \bibnamefont{{\it{et~al.}}} (\bibinfo{collaboration}{CDF Collaboration}),
  \bibinfo{journal}{Phys. Rev. D} \textbf{\bibinfo{volume}{87}},
  \bibinfo{pages}{092002} (\bibinfo{year}{2013}{\natexlab{a}}).

\bibitem[{\citenamefont{Abazov et~al.}(2011{\natexlab{a}})}]{Abazov:2011rq}
\bibinfo{author}{\bibfnamefont{V.~M.} \bibnamefont{Abazov}}
  \bibnamefont{{\it{et~al.}}} (\bibinfo{collaboration}{D0 Collaboration}),
  \bibinfo{journal}{Phys. Rev. D} \textbf{\bibinfo{volume}{84}},
  \bibinfo{pages}{112005} (\bibinfo{year}{2011}{\natexlab{a}}).

\bibitem[{eta()}]{eta}
\bibinfo{note}{{The pseudorapidity $\eta$ is defined as a function of the polar
  angle $\theta$ with respect to the proton beam as \mbox{$\eta =-\ln(\tan
  {{\theta}\over{2}})$}. Positive (negative) $\eta$ corresponds to a particle
  produced in the direction of the incoming proton (antiproton).}}

\bibitem[{\citenamefont{Aaltonen
  et~al.}(2013{\natexlab{b}})}]{2013arXiv1308.1120C}
\bibinfo{author}{\bibfnamefont{T.}~\bibnamefont{Aaltonen}}
  \bibnamefont{{\it{et~al.}}} (\bibinfo{collaboration}{CDF Collaboration})
  (\bibinfo{year}{2013}{\natexlab{b}}), \eprint{arXiv:1308.1120}.

\bibitem[{\citenamefont{Abazov et~al.}(2013{\natexlab{a}})}]{Abazov:2012bfa}
\bibinfo{author}{\bibfnamefont{V.~M.} \bibnamefont{Abazov}}
  \bibnamefont{{\it{et~al.}}} (\bibinfo{collaboration}{D0 Collaboration}),
  \bibinfo{journal}{Phys. Rev. D.} \textbf{\bibinfo{volume}{87}},
  \bibinfo{pages}{011103(R)} (\bibinfo{year}{2013}{\natexlab{a}}).

\bibitem[{\citenamefont{Aad et~al.}(2012)}]{atlas}
\bibinfo{author}{\bibfnamefont{G.}~\bibnamefont{Aad}}
  \bibnamefont{{\it{et~al.}}} (\bibinfo{collaboration}{ATLAS Collaboration}),
  \bibinfo{journal}{Eur. Phys. J.} \textbf{\bibinfo{volume}{C72}},
  \bibinfo{pages}{2039} (\bibinfo{year}{2012}).

\bibitem[{\citenamefont{Chatrchyan et~al.}(2012)}]{cms}
\bibinfo{author}{\bibfnamefont{S.}~\bibnamefont{Chatrchyan}}
  \bibnamefont{{\it{et~al.}}} (\bibinfo{collaboration}{CMS Collaboration}),
  \bibinfo{journal}{Phys. Lett. B} \textbf{\bibinfo{volume}{717}},
  \bibinfo{pages}{129} (\bibinfo{year}{2012}).

\bibitem[{\citenamefont{Bernreuther and Si}(2012)}]{Bernreuther:2012sx}
\bibinfo{author}{\bibfnamefont{W.}~\bibnamefont{Bernreuther}} \bibnamefont{and}
  \bibinfo{author}{\bibfnamefont{Z.-G.} \bibnamefont{Si}},
  \bibinfo{journal}{Phys. Rev. D} \textbf{\bibinfo{volume}{86}},
  \bibinfo{pages}{034026} (\bibinfo{year}{2012}).

\bibitem[{\citenamefont{Abazov et~al.}(2006)}]{run2det}
\bibinfo{author}{\bibfnamefont{V.~M.} \bibnamefont{Abazov}}
  \bibnamefont{{\it{et~al.}}} (\bibinfo{collaboration}{D0 Collaboration}),
  \bibinfo{journal}{Nucl. Instrum. Methods Phys. Res. A}
  \textbf{\bibinfo{volume}{565}}, \bibinfo{pages}{463} (\bibinfo{year}{2006}).

\bibitem[{\citenamefont{Angstadt et~al.}(2010)}]{Angstadt2010}
\bibinfo{author}{\bibfnamefont{R.}~\bibnamefont{Angstadt}}
  \bibnamefont{{\it{et~al.}}} (\bibinfo{collaboration}{D0 Collaboration}),
  \bibinfo{journal}{Nucl. Instrum. Methods Phys. Res. A}
  \textbf{\bibinfo{volume}{622}}, \bibinfo{pages}{298} (\bibinfo{year}{2010}).

\bibitem[{\citenamefont{Abolins et~al.}(2008)}]{Abolins2008}
\bibinfo{author}{\bibfnamefont{M.}~\bibnamefont{Abolins}}
  \bibnamefont{{\it{et~al.}}}, \bibinfo{journal}{Nucl. Instrum. Methods Phys.
  Res. A} \textbf{\bibinfo{volume}{A584}}, \bibinfo{pages}{75}
  (\bibinfo{year}{2008}).

\bibitem[{\citenamefont{Abachi et~al.}(1994)}]{run1det}
\bibinfo{author}{\bibfnamefont{S.}~\bibnamefont{Abachi}}
  \bibnamefont{{\it{et~al.}}} (\bibinfo{collaboration}{D0 Collaboration}),
  \bibinfo{journal}{Nucl. Instrum. Methods Phys. Res. A}
  \textbf{\bibinfo{volume}{338}}, \bibinfo{pages}{185} (\bibinfo{year}{1994}).

\bibitem[{\citenamefont{Abazov et~al.}(2005)}]{run2muon}
\bibinfo{author}{\bibfnamefont{V.~M.} \bibnamefont{Abazov}}
  \bibnamefont{{\it{et~al.}}} (\bibinfo{collaboration}{D0 Collaboration}),
  \bibinfo{journal}{Nucl. Instrum. Methods Phys. Res. A}
  \textbf{\bibinfo{volume}{552}}, \bibinfo{pages}{372} (\bibinfo{year}{2005}).

\bibitem[{\citenamefont{Abazov et~al.}(2013{\natexlab{b}})}]{muonID}
\bibinfo{author}{\bibfnamefont{V.~M.} \bibnamefont{Abazov}}
  \bibnamefont{{\it{et~al.}}} (\bibinfo{collaboration}{D0 Collaboration}),
  \bibinfo{journal}{submitted to Nucl. Instrum. Methods Phys. Res. A,
  arXiv:1307.5202 [hep-ex]}  (\bibinfo{year}{2013}{\natexlab{b}}).

\bibitem[{\citenamefont{Blazey et~al.}(2000)}]{Blazey:2000qt}
\bibinfo{author}{\bibfnamefont{G.~C.} \bibnamefont{Blazey}}
  \bibnamefont{{\it{et~al.}}}, \bibinfo{journal}{arXiv:hep-ex/0005012}
  (\bibinfo{year}{2000}).

\bibitem[{\citenamefont{Abazov et~al.}(2010{\natexlab{a}})}]{Abazov:2010ab}
\bibinfo{author}{\bibfnamefont{V.~M.} \bibnamefont{Abazov}}
  \bibnamefont{{\it{et~al.}}} (\bibinfo{collaboration}{D0 Collaboration}),
  \bibinfo{journal}{Nucl. Instrum. Methods Phys. Res. A}
  \textbf{\bibinfo{volume}{620}}, \bibinfo{pages}{490}
  (\bibinfo{year}{2010}{\natexlab{a}}).

\bibitem[{\citenamefont{Brun and Carminati}(unpublished)}]{geant}
\bibinfo{author}{\bibfnamefont{R.}~\bibnamefont{Brun}} \bibnamefont{and}
  \bibinfo{author}{\bibfnamefont{F.}~\bibnamefont{Carminati}},
  \bibinfo{journal}{CERN Program Library Long Writeup W5013 (1993)}
  (\bibinfo{year}{unpublished}).

\bibitem[{\citenamefont{Frixione and Webber}(2002)}]{Frixione:2002ik}
\bibinfo{author}{\bibfnamefont{S.}~\bibnamefont{Frixione}} \bibnamefont{and}
  \bibinfo{author}{\bibfnamefont{B.~R.} \bibnamefont{Webber}},
  \bibinfo{journal}{J. High Energy Phys.} \textbf{\bibinfo{volume}{06}}
  (\bibinfo{year}{2002}), \bibinfo{note}{029}.

\bibitem[{\citenamefont{Frixione and Webber}(2008)}]{Frixione:2008ym}
\bibinfo{author}{\bibfnamefont{S.}~\bibnamefont{Frixione}} \bibnamefont{and}
  \bibinfo{author}{\bibfnamefont{B.~R.} \bibnamefont{Webber}},
  \bibinfo{journal}{arXiv:0812.0770}  (\bibinfo{year}{2008}).

\bibitem[{\citenamefont{Corcella et~al.}(2001)}]{Corcella:2000bw}
\bibinfo{author}{\bibfnamefont{G.}~\bibnamefont{Corcella}}
  \bibnamefont{{\it{et~al.}}}, \bibinfo{journal}{J. High Energy Phys.}
  \textbf{\bibinfo{volume}{01}} (\bibinfo{year}{2001}), \bibinfo{note}{010}.

\bibitem[{\citenamefont{Mangano et~al.}(2003)}]{Mangano2003}
\bibinfo{author}{\bibfnamefont{M.~L.} \bibnamefont{Mangano}}
  \bibnamefont{{\it{et~al.}}}, \bibinfo{journal}{J. High Energy Phys.}
  \textbf{\bibinfo{volume}{07}} (\bibinfo{year}{2003}), \bibinfo{note}{001}.

\bibitem[{\citenamefont{Sjostrand et~al.}(2006)\citenamefont{Sjostrand, Mrenna,
  and Skands}}]{Sjostrand2006}
\bibinfo{author}{\bibfnamefont{T.}~\bibnamefont{Sjostrand}},
  \bibinfo{author}{\bibfnamefont{S.}~\bibnamefont{Mrenna}}, \bibnamefont{and}
  \bibinfo{author}{\bibfnamefont{P.~Z.} \bibnamefont{Skands}},
  \bibinfo{journal}{J. High Energy Phys.} \textbf{\bibinfo{volume}{05}}
  (\bibinfo{year}{2006}), \bibinfo{note}{026}.

\bibitem[{\citenamefont{Affolder et~al.}(2002)}]{Affolder2002}
\bibinfo{author}{\bibfnamefont{T.}~\bibnamefont{Affolder}}
  \bibnamefont{{\it{et~al.}}} (\bibinfo{collaboration}{CDF Collaboration}),
  \bibinfo{journal}{Phys. Rev. D} \textbf{\bibinfo{volume}{65}},
  \bibinfo{pages}{092002} (\bibinfo{year}{2002}).

\bibitem[{\citenamefont{Nadolsky et~al.}(2008)}]{Nadolsky:2008zw}
\bibinfo{author}{\bibfnamefont{P.~M.} \bibnamefont{Nadolsky}}
  \bibnamefont{{\it{et~al.}}}, \bibinfo{journal}{Phys. Rev. D}
  \textbf{\bibinfo{volume}{78}}, \bibinfo{pages}{013004}
  (\bibinfo{year}{2008}).

\bibitem[{\citenamefont{Gavin et~al.}(2011)\citenamefont{Gavin, Li, Petriello,
  and Quackenbush}}]{Gavin:2010az}
\bibinfo{author}{\bibfnamefont{R.}~\bibnamefont{Gavin}},
  \bibinfo{author}{\bibfnamefont{Y.}~\bibnamefont{Li}},
  \bibinfo{author}{\bibfnamefont{F.}~\bibnamefont{Petriello}},
  \bibnamefont{and}
  \bibinfo{author}{\bibfnamefont{S.}~\bibnamefont{Quackenbush}},
  \bibinfo{journal}{Comput. Phys. Commun.} \textbf{\bibinfo{volume}{182}},
  \bibinfo{pages}{2388} (\bibinfo{year}{2011}).

\bibitem[{\citenamefont{Ellis}(2006)}]{Ellis:2006ar}
\bibinfo{author}{\bibfnamefont{R.~K.} \bibnamefont{Ellis}},
  \bibinfo{journal}{Nucl. Phys. Proc. Suppl.} \textbf{\bibinfo{volume}{160}},
  \bibinfo{pages}{170} (\bibinfo{year}{2006}).

\bibitem[{\citenamefont{Abazov et~al.}(2010{\natexlab{b}})}]{Abazov:2010kn}
\bibinfo{author}{\bibfnamefont{V.~M.} \bibnamefont{Abazov}}
  \bibnamefont{{\it{et~al.}}} (\bibinfo{collaboration}{D0 Collaboration}),
  \bibinfo{journal}{Phys. Lett. B} \textbf{\bibinfo{volume}{693}},
  \bibinfo{pages}{522} (\bibinfo{year}{2010}{\natexlab{b}}).

\bibitem[{\citenamefont{Sjostrand et~al.}(2008)\citenamefont{Sjostrand, Mrenna,
  and Skands}}]{Sjostrand2008}
\bibinfo{author}{\bibfnamefont{T.}~\bibnamefont{Sjostrand}},
  \bibinfo{author}{\bibfnamefont{S.}~\bibnamefont{Mrenna}}, \bibnamefont{and}
  \bibinfo{author}{\bibfnamefont{P.~Z.} \bibnamefont{Skands}},
  \bibinfo{journal}{Comput. Phys. Commun.} \textbf{\bibinfo{volume}{178}},
  \bibinfo{pages}{852} (\bibinfo{year}{2008}).

\bibitem[{\citenamefont{Falkowski et~al.}(2012)\citenamefont{Falkowski,
  Mangano, Martin, Perez, and Winter}}]{Falkowski:2012cu}
\bibinfo{author}{\bibfnamefont{A.}~\bibnamefont{Falkowski}},
  \bibinfo{author}{\bibfnamefont{M.~L.} \bibnamefont{Mangano}},
  \bibinfo{author}{\bibfnamefont{A.}~\bibnamefont{Martin}},
  \bibinfo{author}{\bibfnamefont{G.}~\bibnamefont{Perez}}, \bibnamefont{and}
  \bibinfo{author}{\bibfnamefont{J.}~\bibnamefont{Winter}}
  (\bibinfo{year}{2012}), \eprint{arXiv:1212.4003}.

\bibitem[{\citenamefont{Abazov et~al.}(2011{\natexlab{b}})}]{Abazov:2011cq}
\bibinfo{author}{\bibfnamefont{V.~M.} \bibnamefont{Abazov}}
  \bibnamefont{{\it{et~al.}}} (\bibinfo{collaboration}{D0 Collaboration}),
  \bibinfo{journal}{Phys. Lett. B} \textbf{\bibinfo{volume}{704}},
  \bibinfo{pages}{403} (\bibinfo{year}{2011}{\natexlab{b}}).

\bibitem[{\citenamefont{Abazov et~al.}(2007)}]{Abazov:2007kg}
\bibinfo{author}{\bibfnamefont{V.}~\bibnamefont{Abazov}}
  \bibnamefont{{\it{et~al.}}} (\bibinfo{collaboration}{D0 Collaboration}),
  \bibinfo{journal}{Phys. Rev.} \textbf{\bibinfo{volume}{D76}},
  \bibinfo{pages}{092007} (\bibinfo{year}{2007}).

\bibitem[{\citenamefont{Moch and Uwer}(2008)}]{PhysRevD.78.034003}
\bibinfo{author}{\bibfnamefont{S.}~\bibnamefont{Moch}} \bibnamefont{and}
  \bibinfo{author}{\bibfnamefont{P.}~\bibnamefont{Uwer}},
  \bibinfo{journal}{Phys. Rev. D} \textbf{\bibinfo{volume}{78}},
  \bibinfo{pages}{034003} (\bibinfo{year}{2008}).

\bibitem[{\citenamefont{Andeen et~al.}(2007)}]{Andeen:2007zc}
\bibinfo{author}{\bibfnamefont{T.}~\bibnamefont{Andeen}}
  \bibnamefont{{\it{et~al.}}} (\bibinfo{collaboration}{D0 Collaboration}),
  \bibinfo{journal}{FERMILAB-TM-2365}  (\bibinfo{year}{2007}).

\bibitem[{\citenamefont{Valassi}(2003)}]{Valassi2003391}
\bibinfo{author}{\bibfnamefont{A.}~\bibnamefont{Valassi}},
  \bibinfo{journal}{Nucl. Instrum. Methods Phys. Res. A}
  \textbf{\bibinfo{volume}{500}}, \bibinfo{pages}{391 } (\bibinfo{year}{2003}).

\bibitem[{\citenamefont{Lyons et~al.}(1988)\citenamefont{Lyons, Gibaut, and
  Clifford}}]{Lyons1988110}
\bibinfo{author}{\bibfnamefont{L.}~\bibnamefont{Lyons}},
  \bibinfo{author}{\bibfnamefont{D.}~\bibnamefont{Gibaut}}, \bibnamefont{and}
  \bibinfo{author}{\bibfnamefont{P.}~\bibnamefont{Clifford}},
  \bibinfo{journal}{Nucl. Instrum. Methods Phys. Res. A}
  \textbf{\bibinfo{volume}{270}}, \bibinfo{pages}{110 } (\bibinfo{year}{1988}).

\end{thebibliography}


\end{document}